\documentclass[11pt,epsf]{article}
\DeclareMathAlphabet{\scr}{U}{rsfs}{m}{n}
\usepackage{latexsym,color}
\usepackage{epsfig}
\usepackage[mathscr]{eucal}
\usepackage{amsfonts}
\usepackage{amscd}
\usepackage{cite}
\usepackage{array}
\usepackage{amssymb}
\usepackage{colordvi}
\usepackage[centertags]{amsmath}
\usepackage{enumerate}
\usepackage{graphicx}
\usepackage{booktabs}
\usepackage{theorem}
\usepackage[footnotesize]{caption}
\usepackage{soul}
\newcommand{\newc}{\newcommand}
\newc{\be}{\begin{equation}}
\newc{\ee}{\end{equation}}
\newc{\bea}{\begin{eqnarray}}
\newc{\eea}{\end{eqnarray}}
\newc{\ol}{\overline}
\newc{\wt}{\widetilde}
\newc{\bs}{\boldsymbol}
\newc{\m}{\mathcal}
\newc{\la}{\langle}
\newc{\ra}{\rangle}

\newcommand{\beq}{\begin{eqnarray}}
\newcommand{\eeq}{\end{eqnarray}}
\newcommand{\s}{\smallskip}

\newcommand{\bc}{\begin{center}}
\newcommand{\ec}{\end{center}}

\newcommand{\ba}{\begin{array}}
\newcommand{\ea}{\end{array}}

\begin{document}

\title{
\vspace*{-3cm}
\phantom{h} \hfill\mbox{\small IPPP/17/37}\\[-1.1cm]
\phantom{h} \hfill\mbox{\small KA-TP-22-2017}\\[-1.1cm]
\phantom{h} \hfill\mbox{\small PSI-PR-17-08}
%\vspace*{1cm}
\\[1cm]
%\vspace{13mm}
\textbf{Higgs Pair Production at NLO QCD \\ for CP-violating
  Higgs Sectors}}

\date{}
\author{
R.~Gr\"ober$^{1\,}$\footnote{E-mail: \texttt{ramona.groeber@durham.ac.uk}},
M. M\"{u}hlleitner$^{2\,}$\footnote{E-mail:
  \texttt{milada.muehlleitner@kit.edu}} $\;$and
M.~Spira$^{3\,}$\footnote{E-mail: \texttt{michael.spira@psi.ch}}\\[9mm]
{\small\it
$^1$ Institute of Particle Physics Phenomenology, Physics Department, Durham University, Durham DH1 3LE, UK}\\[3mm] 
{\small\it
$^2$Institute for Theoretical Physics, Karlsruhe Institute of Technology,} \\
{\small\it 76128 Karlsruhe, Germany}\\[3mm]
{\small\it
$^3$ Paul Scherrer Institute, CH-5323 Villigen PSI, Switzerland}}

\maketitle

\begin{abstract}
\noindent
Higgs pair production through gluon fusion is an important process at
the LHC to test the dynamics underlying electroweak symmetry
breaking. Higgs sectors beyond the Standard Model (SM) can
substantially modify this cross section through novel couplings not
present in the SM or the on-shell production of new heavy Higgs bosons
that subsequently decay into Higgs pairs. CP violation in the Higgs
sector is important for the explanation of the observed
matter-antimatter asymmetry through electroweak baryogenesis. In this
work we compute the next-to-leading order (NLO) QCD corrections in the
heavy top quark limit, including the effects of CP violation in
the Higgs sector. We choose the effective theory (EFT) approach, which
provides a rather model-independent way to explore New Physics (NP) effects
by adding dimension-6 operators, both CP-conserving and CP-violating
ones, to the SM Lagrangian. Furthermore, we perform the 
computation within a specific UV-complete model and choose as
benchmark model the general 2-Higgs-Doublet Model with CP violation,
the C2HDM. Depending on the dimension-6 coefficients, the relative NLO QCD
corrections are affected by several per cent through the new
CP-violating operators. This is also 
the case for SM-like Higgs pair production in the C2HDM, while the
relative QCD corrections in the production of heavier C2HDM Higgs boson pairs
deviate more strongly from the SM case. The absolute cross sections
both in the EFT and the C2HDM can be modified by more than an order of
magnitude. In particular, in the C2HDM the resonant production of Higgs pairs can by
far exceed the SM cross section.
\end{abstract}
\thispagestyle{empty}
\vfill
\newpage
\setcounter{page}{1}

%%%%%%%%%%%%%%%%%%%%%%%%%%%%%%%%%%%%%%%%%%%%%%%%%%%%%%%
\section{Introduction}
With the discovery of the Higgs boson \cite{Aad:2012tfa,Chatrchyan:2012ufa}, the
Standard Model (SM) is structurally complete. While the Higgs boson behaves
very SM-like, the open questions that cannot be answered within the
SM, call for New Physics (NP) extensions. In view of the lack of direct
discoveries of particles predicted by extensions beyond the SM (BSM), the
precise investigation of the Higgs sector plays an important role \cite{Englert:2014uua}. 
The Higgs self-couplings determine the shape of the Higgs
potential. Although the Higgs couplings to the SM particles are very
SM-like, the Higgs self-couplings can still deviate substantially from
their SM values
\cite{DiLuzio:2017tfn}. The trilinear
Higgs self-coupling is directly accessible in Higgs pair production
\cite{Dawson:1998py,Djouadi:1999gv,Djouadi:1999rca,Muhlleitner:2000jj}. At the LHC
gluon fusion into Higgs pairs 
provides the largest Higgs pair production cross section
\cite{Glover:1987nx,Plehn:1996wb,Baglio:2012np}. With a value of 32.91
fb at NLO QCD including the full top quark mass dependence for a Higgs
mass of 125 GeV and a c.m.~energy of $\sqrt{s}=14$~TeV 
\cite{Borowka:2016ehy,Borowka:2016ypz,Heinrich:2017kxx} 
%(+13.8-12.8\%, 13 TeV; +13.6-12.6\%, 14 TeV 2.paper, 173 GeV, 125 GeV)
this process is experimentally challenging. In BSM models,
however, Higgs pair production can be significantly enhanced, see {\it
  e.g.}~\cite{Djouadi:1999rca,Ellwanger:2013ova,Nhung:2013lpa,No:2013wsa,Heng:2013cya,Bhattacherjee:2014bca,King:2014xwa,Chen:2014ask,Martin-Lozano:2015dja,Wu:2015nba,He:2015spf,Dawson:2015haa,Batell:2015koa,Zhang:2015mnh,Bojarski:2015kra,Grober:2015cwa,Grober:2016wmf,Kanemura:2016tan,He:2016sqr,Krause:2016xku,Baglio:2016bop,Huang:2017jws,Nakamura:2017irk}. \s

Higgs pair production through gluon fusion is mediated by top and bottom
quark triangle and box diagrams already at leading order (LO).
The next-to-leading order (NLO) QCD corrections are important and have first 
been obtained in the limit of large top quark masses
\cite{Dawson:1998py}. Top quark mass effects are important and have been
analysed in
\cite{Dawson:2012mk,Grigo:2013rya,Frederix:2014hta,Maltoni:2014eza}
with first results towards a fully differential NLO calculation presented in
\cite{Frederix:2014hta}. In Ref.~\cite{Degrassi:2016vss} analytic
results for the one-particle irreducible contributions to the virtual
NLO QCD corrections were presented including
finite top quark mass effects. Recently, the NLO QCD corrections have been
calculated including the full mass dependence of the top quark in
the loops \cite{Borowka:2016ehy,Borowka:2016ypz,Heinrich:2017kxx}. The
results confirm the relevance of the mass 
effects, in particular for the differential distributions (for former
investigations, see \cite{Baur:2002rb,Gillioz:2012se}). The
next-to-next-to-leading order (NNLO) QCD corrections in the heavy
quark limit have been calculated in
\cite{deFlorian:2013uza,deFlorian:2013jea,Grigo:2014jma}. Results for
differential Higgs pair production at NNLO QCD have been presented in
\cite{deFlorian:2016uhr}. Soft gluon resummation at
next-to-leading logarithmic order within the SCET approach has
been performed in \cite{Shao:2013bz} and extended to the
next-to-next-to-leading logarithmic order in \cite{deFlorian:2015moa},
including also the matching to the NNLO cross section. The NLO QCD
corrections to Higgs pair production in the MSSM in the heavy top mass limit have first
been evaluated in \cite{Dawson:1998py}. More recently, analytic
results for the contributions from one- and two-loop box diagrams involving
top and stop quarks have been obtained in the limit of large loop
particle masses in \cite{Agostini:2016vze}. The NLO QCD corrections to
double Higgs production in the singlet-extended SM have been computed
in \cite{Dawson:2015haa} and those in
the 2-Higgs Doublet Model (2HDM) in \cite{Hespel:2014sla}, both in the
large top mass limit.\s  

A model-independent way to parametrise NP
effects realised at a scale well above the scale of electroweak
symmetry breaking (EWSB), is given by the Effective Field Theory (EFT) approach where
higher-dimensional operators are added to the SM
Lagrangian and lead to modifications of the Higgs boson couplings. 
The impact on Higgs pair production through higher-dimensional
operators was analysed in
\cite{Contino:2012xk,Chen:2014xra,Goertz:2014qta,Edelhaeuser:2015zra,Azatov:2015oxa,Grober:2015cwa,Lu:2015jza,He:2015spf,Kilian:2017nio,deFlorian:2017qfk,Dawson:2017vgm,Corbett:2017ieo}. The
higher-dimensional operators do not only modify the Higgs couplings to
the SM particles but introduce novel couplings not present in the SM, with different
effects on the triangle and box diagrams, too. In
\cite{Grober:2015cwa} we computed the 
NLO QCD corrections to gluon fusion into Higgs pairs including higher
dimensional operators in the large top mass limit. While the new
operators modify the cross section by up to an order of magnitude,
their effect on the relative NLO QCD corrections is only of the order of
several per cent. The NLO QCD corrections to composite Higgs pair
production in models
without and with new heavy fermions have been provided in
\cite{Grober:2016wmf}. Recently, the NNLO QCD corrections in the heavy
loop particle limit have been given in ~\cite{deFlorian:2017qfk} for
the inclusive as well as the 
differential cross section including the relevant dimension-6 operators.
Since the leading order cross section dominantly factorises, the
relative QCD corrections are found to be almost insensitive to the
composite character of the Higgs boson and to the details of the 
heavy fermion spectrum.
\s

In this paper we extend the NLO QCD corrections in the heavy top
quark limit to Higgs sectors including CP violation. In the SM CP
violation is incorporated in the Cabibbo-Kobayashi-Maskawa matrix and
rather small. Models beyond the SM provide additional sources
of CP violation, that can be significant while still being compatible
with the constraints from electric dipole moments (EDMs), see {\it
  e.g.}~\cite{King:2015oxa,Muhlleitner:2017dkd}. CP violation is one
of the three Sakharov conditions 
\cite{Sakharov:1967dj} necessary for baryogenesis. Its discovery in the
Higgs sector provides an immediate proof of physics beyond the SM. We
investigate CP violation both in a more model-independent EFT approach by
adding additional CP-violating dimension-6 operators
\cite{Contino:2013kra,Cao:2016zob} and in a specific benchmark model
given by the CP-violating 2-Higgs-Doublet Model (C2HDM) \cite{Ginzburg:2002wt,Khater:2003wq,ElKaffas:2006gdt,ElKaffas:2007rq,WahabElKaffas:2007xd,Osland:2008aw,Grzadkowski:2009iz,Arhrib:2010ju,Barroso:2012wz,Fontes:2014xva}. \s

The organization of our paper is as follows. In section
\ref{sec:eftcalc} we present the results for the NLO QCD corrections
in the EFT approach including CP-violating dimension-6 operators. The
subsequent section \ref{sec:c2dhmcalc} contains our NLO results in
the CP-violating 2HDM. The numerical analysis is presented in section
\ref{sec:numerical}. In section \ref{sec:concl} we summarise and conclude.

%%%%%%%%%%%%%%%%%%%%%%%%%%%%%%%%%%%%%%%%%%%%%%%%%%%%%%%
\section{Higgs Pair Production in the EFT including CP
  violation \label{sec:eftcalc}}
Before presenting our analytic results for the NLO QCD corrections to
Higgs pair production in the EFT approach including CP violation we
introduce our notation. 

%%%%%%%%%%%%%%%%%%%%%%%%%%%%%%%%%%%%%%%%%%%%%%%%%%%%%%%
\subsection{The EFT including CP violation}
By adding higher-dimensional operators to the SM, NP effects that
appear at scales far above the EWSB scale, can be parametrised in a
model-independent way. In case of a linearly realised $SU(2)_L
\times U(1)_Y$ symmetry the Higgs boson is embedded in an $SU(2)_L$
doublet $H$. The leading BSM effects are then parametrised by
dimension-6 operators. Note, that even though dimension-8 operators
can become more important \cite{Azatov:2015oxa}, the investigation of
the involved kinematic regions is challenging so that we will neglect
them in the following. Adopting the Strongly-Interacting-Light Higgs (SILH) basis the
operators that are relevant for Higgs pair production are given by \cite{Giudice:2007fh},
\beq
\Delta {\cal L}_6^{\text{SILH}} &\supset&
\frac{\bar{c}_H}{2v^2} \partial_\mu (H^\dagger H) \partial^\mu
(H^\dagger H) + \frac{(\bar{c}_u + i\widetilde{\bar{c}}_u)}{v^2} y_t
H^\dagger H \bar{q}_L H^c t_R + h.c. \nonumber \\
&& - \frac{\bar{c}_6}{6 v^2} \frac{3 M_h^2}{v^2} (H^\dagger H)^3
+ \bar{c}_g \frac{g_s^2}{M_W^2} H^\dagger H G_{\mu\nu}^a G^{a\,
  \mu\nu} 
+ \widetilde{\bar{c}}_g \frac{g_s^2}{M_W^2} H^\dagger H G_{\mu\nu}^a \tilde{G}^{a\,
  \mu\nu}
\;, \label{eq:silh}
\eeq 
where $v$ is the vacuum expectation value (VEV) $v\approx 246$~GeV,
$M_h=125.09$~GeV \cite{Aad:2015zhl} the Higgs boson mass, $M_W$ 
the $W$ boson mass, $y_t$ the top Yukawa coupling constant and $g_s$ the
strong coupling constant. The gluon field strength tensor
$G^a_{\mu\nu}$ in terms of the gluon fields $g_\mu^a$ and the $SU(3)$
structure constants $f^{abc}$ is given by
\beq
G^a_{\mu\nu} = \partial_\mu g_\nu^a - \partial_\nu g^a_\mu + g_s
f^{abc} g_\mu^b g_\nu^c \;,
\eeq
and its dual $\tilde{G}^a_{\mu\nu}$ reads
\beq
\tilde{G}^a_{\mu\nu} = \frac{1}{2} \epsilon_{\mu\nu\alpha\beta}
G^{a,\alpha\beta} \;,
\eeq
where $\epsilon_{\mu\nu\alpha\beta}$ it the totally antisymmetric
tensor in four dimensions, normalized to $\epsilon_{0123}=1$. 
The effect of the first three operators in Eq.~(\ref{eq:silh}) is the
modification of the top Yukawa and the trilinear Higgs
self-coupling compared to their SM values. The second operator also 
induces a novel two-Higgs two-fermion coupling \cite{Grober:2010yv}.
The last two operators parametrise effective gluon couplings to one and two Higgs
bosons not mediated by SM quark loops. CP violation is accounted for
by the complex part $i\widetilde{\bar{c}}_u$ of the Yukawa couplings and
the last operator in Eq.~(\ref{eq:silh}) containing the dual gluon
field strength tensor. An estimate of the size of the CP-conserving coefficients
$\bar{c}_H, \bar{c}_u, \bar{c}_6$ and $\bar{c}_g$ and the most
important experimental bounds can be found in
\cite{Contino:2013kra}. 
\s

In case of a non-linearly realised EW symmetry with the physical Higgs
boson $h$ being a singlet of the custodial symmetry and not necessarily
being part of a weak doublet, the non-linear Lagrangian \cite{Contino:2010mh} with the
contributions relevant for Higgs pair production reads 
\beq
\Delta {\cal L}_{\text{non-lin}} \supset &- m_t \bar{t}t \left(c_t
  \frac{h}{v} + c_{tt} \frac{h^2}{2 v^2} \right) - i m_t \bar{t}\gamma_5 t \left(\tilde{c}_t
  \frac{h}{v} + \tilde{c}_{tt} \frac{h^2}{2 v^2} \right) 
-c_3 \, \frac{1}{6} \left( \frac{3 M_h^2}{v} \right) h^3 \nonumber \\ & + \frac{\alpha_s}{\pi} G^{a\, \mu\nu}
G_{\mu\nu}^a \left( c_g \frac{h}{v} + c_{gg}\frac{h^2}{2 v^2}\right) +\frac{\alpha_s}{\pi} G^{a\, \mu\nu}
\tilde{G}_{\mu\nu}^a \left( \tilde{c}_g \frac{h}{v} + \tilde{c}_{gg}\frac{h^2}{2 v^2}
\right) \;,
\eeq
with $\alpha_s = g_s^2/(4\pi)$. Here, the operators with the
coefficients $\tilde{c}_t$, $\tilde{c}_{tt}$, $\tilde{c}_g$ and
$\tilde{c}_{gg}$ account for CP violation.  
A bound on $\tilde{c}_g$ has been given in \cite{Ferreira:2016jea}.
While in the SILH parametrisation the coupling  deviations from the SM are
required to be small, the couplings $c_i$ in the non-linear Lagrangian
can take arbitrary values. The relations between the SILH coefficients
and the non-linear ones can be derived from the SILH Lagrangian in the
unitary gauge after canonical normalization. They read \cite{Azatov:2015oxa}
\beq
c_t &=& 1 - \frac{\bar{c}_H}{2} - \bar{c}_u \;, \quad c_{tt} = -
\frac{1}{2} (\bar{c}_H + 3 \bar{c}_u ) \;, \quad c_3 = 1- \frac{3}{2}
\bar{c}_H + \bar{c}_6 \;, \quad c_g = c_{gg} = \bar{c}_g \left(
  \frac{4 \pi}{\alpha_2} \right) \;, \nonumber\\
\tilde{c}_t &=& - \widetilde{\bar{c}}_u \;, \quad \tilde{c}_{tt} = -
\frac{3}{2} \widetilde{\bar{c}}_u  \;, \quad \tilde{c}_g =
\tilde{c}_{gg} = \widetilde{\bar{c}}_g \left(  \frac{4 \pi}{\alpha_2}
\right) \;, \label{eq:nonlincoeff} 
\eeq
where $\alpha_2 = \sqrt{2} G_F M_W^2/\pi$, with $G_F$ denoting the Fermi
constant. We will give results for the non-linear
parametrisation in the following and summarise the SILH case in 
Appendix~\ref{app:silhggfus}. \s 
  
%%%%%%%%%%%%%%%%%%%%%%%%%%%%%%%%%%%%%%%%%%%%%%%%%%%%%%%%%%%
\subsection{The NLO QCD Corrections in the EFT \label{sec:eftnlocalc}}
\begin{figure}[h]
\begin{minipage}{.25\textwidth}
\includegraphics{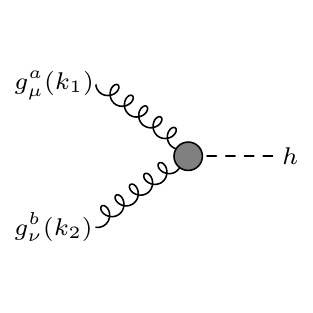}
\end{minipage}
\begin{minipage}{.75\textwidth}
\begin{flushleft}
\beq
i \delta^{ab} \frac{\alpha_s}{\pi v} \Big\{ && \hspace*{-0.6cm} \frac{1}{3}[k_1^\nu k_2^\mu - (k_1 \cdot
k_2) g^{\mu\nu}] \left[ c_t \left(1+\frac{11}{4} \frac{\alpha_s}{\pi} \right)+
  12 c_{g} \right] \nonumber \\ + && \hspace*{-0.6cm} \frac{1}{2}
\epsilon^{\mu\nu\rho\sigma}k_{1\rho}k_{2\sigma} \left[ -\tilde{c}_t
  +8 \tilde{c}_g \right] \Big\}
\nonumber
\eeq
  \end{flushleft}
\end{minipage}\\
\begin{minipage}{.25\textwidth}
\includegraphics{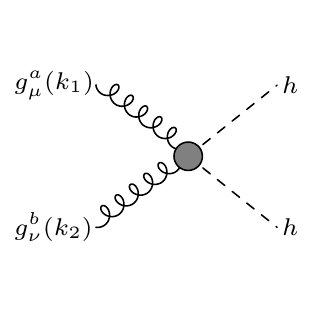}
\end{minipage}
\begin{minipage}{.75\textwidth}
\begin{flushleft}
\beq
i \delta^{ab} \frac{\alpha_s}{\pi v^2} \Big\{ && \hspace*{-0.6cm} \frac{1}{3}[k_1^\nu k_2^\mu - (k_1 \cdot
k_2) g^{\mu\nu}] \left[ (c_{tt}-c_t^2+\tilde{c}_t^2) \left(1+\frac{11}{4}
    \frac{\alpha_s}{\pi} \right) + 12 c_{gg}\right] \nonumber \\ 
+ && \hspace*{-0.6cm} \frac{1}{2}
\epsilon^{\mu\nu\rho\sigma}k_{1\rho}k_{2\sigma}
\left[-\tilde{c}_{tt}+2 c_t \tilde{c}_t +8
  \tilde{c}_{gg}\right] 
\nonumber
\eeq \\
\end{flushleft}
\end{minipage}
\caption{Feynman rules for the effective two-gluon couplings to one (upper)
  and two (lower) CP-violating Higgs bosons in the heavy quark limit,
  including NLO QCD corrections. The four-momenta of the
  gluons, $k_1$ and $k_2$, are taken as both incoming or both
  outgoing. \label{fig:effcpmixedvertices}} 
\end{figure}

Top and bottom quark loops provide the dominant contributions to gluon
fusion into Higgs pairs \cite{Plehn:1996wb}. In the computation of
the NLO QCD corrections in the heavy top quark limit we consistently
neglect the bottom quark loops in the following. Their contribution in
the SM amounts to less than 1\% \cite{Dawson:1998py,Baglio:2012np}. 
For the computation of the QCD corrections to Higgs pair production in
the large top mass limit an effective Lagrangian can be used that is
valid for light Higgs bosons. It contains the Higgs boson interactions
derived in the low-energy limit of small Higgs four-momentum. 
In the case of SM single-Higgs production the $K$-factor derived in this
limit approximates the result obtained with the full mass dependence
to better than 5\%
\cite{Graudenz:1992pv,Spira:1993bb,Spira:1995rr,Kramer:1996iq,Harlander:2005rq}. In
Higgs pair production, the low-energy approach works less well and
induces an uncertainty of about 15\% in the $K$-factor
\cite{Borowka:2016ehy,Borowka:2016ypz,Heinrich:2017kxx}.
Note, that the top mass effects on the $K$-factor for models including
higher-dimensional operators can also be expected to be of order
10--20\%, as the NLO corrections are dominated by soft and collinear
gluon effects. The Lagrangian with the required effective Higgs
couplings to gluons and quarks can be derived from \cite{Dawson:1998py} as
\beq
{\cal L}_{\text{eff}} = \frac{\alpha_s}{\pi} G^{a\mu\nu}
G_{\mu\nu}^a && \hspace*{-0.6cm} \left\{ \frac{h}{v} \left[\frac{c_t}{12} \left( 1 +
      \frac{11}{4} \frac{\alpha_s}{\pi} \right) + c_g
  \right] \right. \nonumber \\
&& \hspace*{-0.6cm} \left. + \frac{h^2}{v^2} \left[
    \frac{-c_t^2+c_{tt}+\tilde{c}_t^2}{24} \left( 1 + 
      \frac{11}{4} \frac{\alpha_s}{\pi} \right) + \frac{c_{gg}}{2}
  \right] \right\} \nonumber \\
+ \frac{\alpha_s}{\pi} G^{a\mu\nu}
\tilde{G}_{\mu\nu}^a &&  \hspace*{-0.6cm} \left\{ \frac{h}{v}
  \left[-\frac{\tilde{c}_t}{8} + \tilde{c}_g
  \right] + \frac{h^2}{v^2} \left[
    \frac{c_t \tilde{c}_t}{8} - \frac{\tilde{c}_{tt}}{16} + \frac{\tilde{c}_{gg}}{2}
  \right] \right\} \;.
\eeq
The factor $(1+11/4 \, \alpha_s/\pi)$ arises from the matching of the effective 
to the full theory at NLO QCD. Note, that neither the effective couplings to
gluons nor the purely CP-odd contributions to the Lagrangian receive this factor. 
The Feynman rules for the effective
couplings between one or two Higgs bosons and two gluons, obtained from
this Lagrangian based on the low-energy theorems
\cite{Ellis:1975ap,Shifman:1979eb,Kniehl:1995tn} are summarised in
Fig.~\ref{fig:effcpmixedvertices}. \s

\begin{figure}[t]
\begin{minipage}{0.33\textwidth}
 \includegraphics[width=\textwidth]{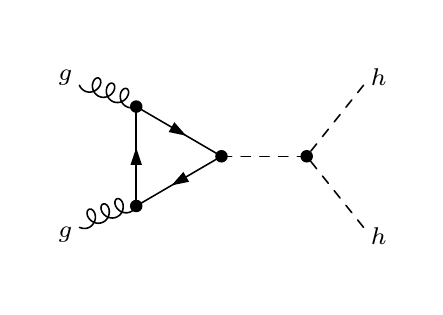}
\end{minipage}
\begin{minipage}{0.33\textwidth}
 \includegraphics[width=\textwidth]{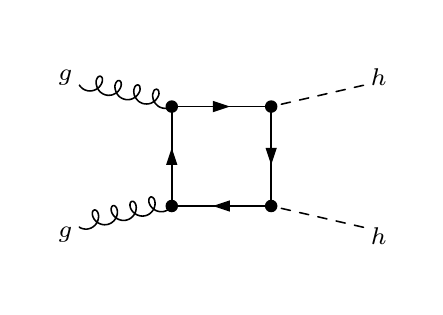}
\end{minipage}
\begin{minipage}{0.33\textwidth}
 \includegraphics[width=\textwidth]{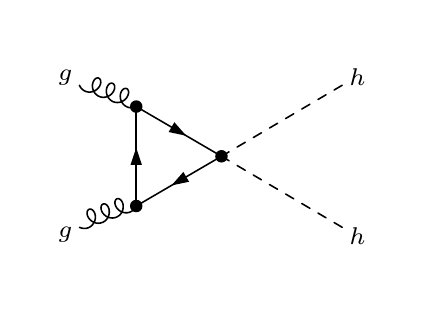}
\end{minipage}\\
 \begin{minipage}{0.49\textwidth}
\begin{flushright}
 \includegraphics[width=0.66\textwidth]{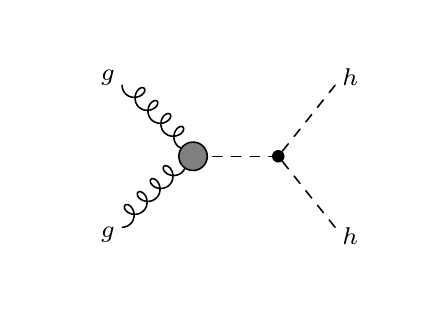}
\end{flushright}
\end{minipage}
\begin{minipage}{0.49\textwidth}
\begin{flushleft}
 \includegraphics[width=0.66\textwidth]{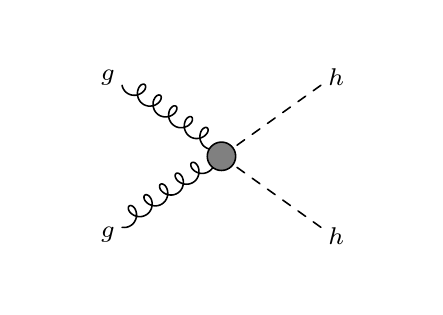}
\end{flushleft}
\end{minipage}

\caption{Generic diagrams contributing to Higgs pair production in
  gluon fusion at LO. \label{fig:logghhdiagrams}} 
\end{figure}
Figure~\ref{fig:logghhdiagrams} shows the generic diagrams that
contribute to Higgs pair production through gluon fusion. Applying the
effective Feynman rules of Fig.~\ref{fig:effcpmixedvertices} results
in the LO partonic cross section, which can be written as,
\beq
\hat{\sigma}_{\text{LO}} (gg \to hh) = \int_{\hat{t}_-}^{\hat{t}_+} 
d\hat{t} \, \frac{G_F^2 \alpha_s^2(\mu_R)}{512 (2\pi)^3} \,
\left[
\left| C_\Delta F_1 + F_2 \right|^2   + |G_1|^2 +
\left| C_\Delta \tilde{F}_1 + \tilde{F}_2 \right|^2   + |\tilde{G}_1|^2 
 \right] \;, \label{eq:losigma} 
\eeq
where $\mu_R$ denotes the renormalisation scale. The Mandelstam
variables read 
\beq
\hat{s} &=& Q^2 \;, \qquad 
\hat{t} = M_h^2 - \frac{Q^2(1-\beta \cos \theta)}{2}  \qquad
\mbox{and} \qquad
\hat{u} = M_h^2 - \frac{Q^2(1+\beta \cos \theta)}{2} \;,
\eeq
in terms of the scattering angle $\theta$ in the partonic
center-of-mass (c.m.)~system
with the invariant Higgs pair mass $Q$ and the relative velocity
\beq
\beta = \sqrt{1-\frac{4M_h^2}{Q^2}}\;.
\eeq
The integration limits are given by $\cos\theta = \pm 1$, {\it i.e.}
\beq
\hat{t}_\pm = M_h^2 -\frac{Q^2(1 \mp \beta)}{2} \;.
\eeq
The $F_{1,2}$, $\tilde{F}_{1,2}$, $G_1$ and $\tilde{G}_1$ summarise
the various form factor contributions with their corresponding
coupling coefficients. They can be cast into the form
\beq
F_1 &=& c_t F_\Delta^e + \frac{2}{3} c_\Delta \nonumber \\
F_2 &=& c_t^2 F^e_\Box + \tilde{c}_t^2 F^o_\Box
+c_{tt} F^e_\Delta - \frac{2}{3} c_\Box \nonumber \\
G_1 &=& c_t^2 G_\Box^e + \tilde{c}_t^2 G_\Box^o \nonumber \\
\tilde{F}_1 &=& \tilde{c}_t F_\Delta^o + \tilde{c}_\Delta \nonumber \\
\tilde{F}_2 &=& 2 c_t \tilde{c}_t F^m_\Box + \tilde{c}_{tt} F_\Delta^o
-\tilde{c}_\Box \nonumber \\
\tilde{G}_1 &=& 2 c_t \tilde{c}_t G^m_\Box \;.
\eeq
The form factors contain the full mass dependence and have been given
in \cite{Plehn:1996wb}. 
The triangle form factors for the projection on the CP-even and
CP-odd Higgs component, $F^e_\Delta$ and $F^o_\Delta$, are given by $F_\Delta$
in appendix A1 and by $F^A_\Delta$ in A2 of \cite{Plehn:1996wb},
respectively. The box form factors corresponding to the
spin-0 gluon-gluon couplings, $F_{\Box}^e$, $F_{\Box}^o$ and
$F_\Box^{m}$, projecting on a purely CP-even, purely CP-odd and a
CP-mixed final state Higgs pair, respectively, are given by $F_\Box$
of appendix A1, A3 and A2. Finally, the CP-even, CP-odd and
CP-mixed box form factors $G_{\Box}^e$, $G_{\Box}^o$ and $G_\Box^{m}$
corresponding to the spin-2 gluon-gluon couplings are the $G_\Box$
form factors of appendix A1, A3 and A2, respectively. 
In the heavy quark limit the form factors read
%\beq
%&& F^{e,\text{lim}}_\Delta = \frac{2}{3} \;, \; F^{o,\text{lim}}_\Delta =
%1 \;, \; F_\Box^{e,\text{lim}} = - F_\Box^{o,\text{lim}} = -\frac{2}{3} \;, \;
%F_\Box^{m,\text{lim}} = -1 \;, \nonumber \\
%&& G_\Box^{e,\text{lim}} = G_\Box^{o,\text{lim}} = G_\Box^{m,\text{lim}}
%= 0 \;.
%\eeq
\beq
F^{e}_\Delta \to \frac{2}{3} \;, \; F^{o}_\Delta \to 
1 \;, \; -F_\Box^{e} , \, F_\Box^{o} \to \frac{2}{3} \;, \;
F_\Box^{m} \to -1 \;, \;
G_\Box^{e} ,\, G_\Box^{o} ,\, G_\Box^{m} \to 0 \;.
\eeq
The introduced abbreviations are 
\beq
C_\Delta &\equiv& \lambda_{hhh} \frac{M_Z^2}{Q^2-M_h^2+i M_h \Gamma_h}
\\
c_\Delta &\equiv& 12 c_{g} \;,\; c_\Box \equiv - 12 c_{gg} \;, \;
\tilde{c}_\Delta \equiv - 8 \tilde{c}_g \quad \mbox{and} \quad
\tilde{c}_\Box = 8 \tilde{c}_{gg}  \;.
\eeq
The trilinear self-coupling $\lambda_{hhh}$, corresponding to the SM
value modified by $c_3$, is given by
\beq
\lambda_{hhh} = \frac{3 M_h^2 c_3}{M_Z^2} \;.
\eeq
The first terms in $F_1$, $F_2$ and $G_1$, respectively, are the SM contributions
modified by the rescaling $c_t$ of the Yukawa coupling and $c_3$ of
the Higgs self-coupling (contained in $C_\Delta$). The contributions
proportional to $c_\Delta$, $c_\Box$, $\tilde{c}_\Delta$ and
$\tilde{c}_\Box$ originate from the effective two-gluon couplings to one
and two Higgs bosons. The novel 2-Higgs-2-fermion couplings induce the
terms coming with $c_{tt}$ and $\tilde{c}_{tt}$. The form factor
contributions proportional to $\tilde{c}_t$, respectively
$\tilde{c}_t^2$, $\tilde{c}_{tt}$ and the ones coming with
$\tilde{c}_\Delta$ and $\tilde{c}_\Box$ are the new contributions due
to the admission of CP violation. We recover the following
limiting cases for the production of a Higgs pair 
\beq
\begin{array}{lll}
\mbox{SM-like} &:& c_t = c_3 = 1 \;,\;  c_{tt} = c_\Delta = c_\Box =
              \tilde{c}_t = \tilde{c}_{tt} = \tilde{c}_\Delta =
              \tilde{c}_\Box = 0 \\
\mbox{purely CP-even} &:& \tilde{c}_t = \tilde{c}_{tt} =
                          \tilde{c}_\Delta = \tilde{c}_\Box = 0 \\
\mbox{purely CP-odd} &:& c_t = c_{tt} =
                          c_\Delta = c_\Box = 0 \;.
\end{array}
\eeq

\begin{figure}
\begin{minipage}{0.33\textwidth}
\includegraphics[width=\textwidth]{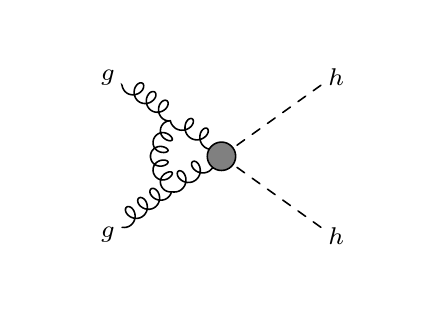}
\end{minipage}
\begin{minipage}{0.33\textwidth}
\includegraphics[width=\textwidth]{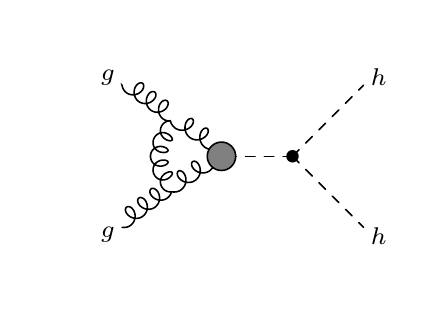}
\end{minipage}
\begin{minipage}{0.33\textwidth}
\includegraphics[width=\textwidth]{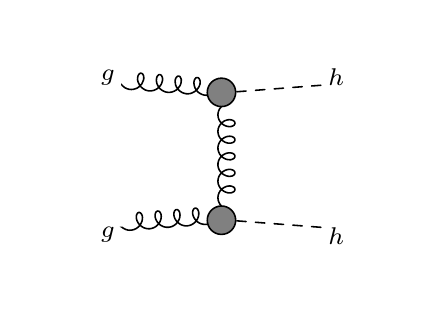}
\end{minipage}\\
\begin{minipage}{0.33\textwidth}
\includegraphics[width=\textwidth]{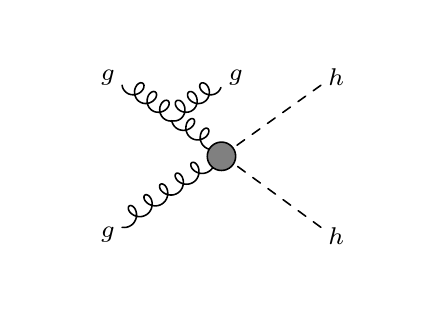}
\end{minipage}
\begin{minipage}{0.33\textwidth}
\includegraphics[width=\textwidth]{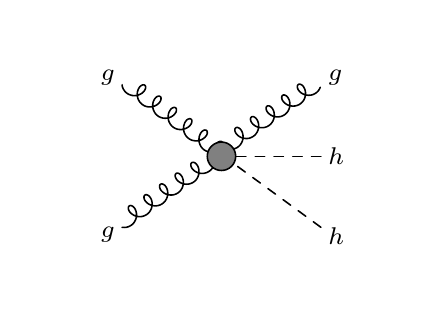}
\end{minipage}
\begin{minipage}{0.33\textwidth}
\includegraphics[width=\textwidth]{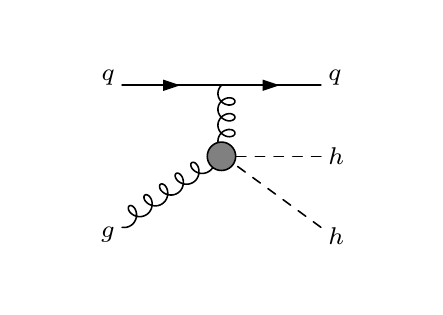}
\end{minipage}
\caption{Sample effective diagrams contributing to the virtual
  (upper) and the real (lower) corrections to gluon fusion into Higgs
  pairs. \label{fig:nlodiags}} 
\end{figure}
The NLO QCD corrections to gluon fusion into Higgs pairs, composed of the
virtual and the real corrections, are obtained
with the help of the effective couplings defined in
Fig.~\ref{fig:effcpmixedvertices}. Sample diagrams are depicted in
Fig.~\ref{fig:nlodiags}. Applying dimensional regularization in
$d=4-2\epsilon$ dimensions, the involved ultraviolet and infrared
divergences appear as poles in $\epsilon$. We renormalise the strong
coupling constant in the $\overline{\mbox{MS}}$ scheme with 
five active flavours, {\it i.e.}~with the top quark decoupled from the
running of $\alpha_s$, in order to cancel the ultraviolet divergences.
The sum of the virtual and real corrections cancels the infrared
divergences. The remaining collinear initial state singularities are
absorbed into the NLO parton densities. These are defined in the
$\overline{\mbox{MS}}$ scheme with five light quark flavours. The
finite hadronic NLO cross section can then be cast into the form
\beq
\sigma_{\text{NLO}} (pp \to hh + X) = \sigma_{\text{LO}} +
\Delta \sigma_{\text{virt}} + \Delta \sigma_{gg} + \Delta \sigma_{gq}
+ \Delta \sigma_{q\bar{q}} \;. \label{eq:nlocxn}
\eeq
The individual contributions of Eq.~(\ref{eq:nlocxn}) read
\beq
\sigma_{\text{LO}} &=& \int_{\tau_0}^1 d\tau \, \frac{d {\cal
    L}^{gg}}{d\tau} \, \hat{\sigma}_{\text{LO}} (Q^2 = \tau
s) \label{eq:contrib0} \nonumber\\
\Delta \sigma_{\text{virt}} &=& \frac{\alpha_s (\mu_R)}{\pi} \int_{\tau_0}^1
d\tau \, \frac{d {\cal L}^{gg}}{d\tau} \, \hat{\sigma}_{\text{LO}} (Q^2 =
\tau s) \, C \label{eq:contrib1} \nonumber \\
\Delta \sigma_{gg} &=& \frac{\alpha_s (\mu_R)}{\pi} \int_{\tau_0}^1
d\tau \, \frac{d {\cal L}^{gg}}{d\tau} 
\int_{\tau_0/\tau}^1 \frac{dz}{z} \, \hat{\sigma}_{\text{LO}} (Q^2 =
z\tau s) \left\{ -z P_{gg} (z) \log \frac{\mu_F^2}{\tau s} \right. \nonumber \\
&& \hspace*{2cm} \left. -\frac{11}{2} (1-z)^3 + 6[1+z^4+(1-z)^4] \left(
    \frac{\log (1-z)}{1-z} \right)_+ \right\} \label{eq:contrib2}
\nonumber \\
\Delta \sigma_{gq} &=& \frac{\alpha_s (\mu_R)}{\pi} \int_{\tau_0}^1
d\tau \, \sum_{q,\bar{q}} \frac{d {\cal L}^{gq}}{d\tau} 
\int_{\tau_0/\tau}^1 \frac{dz}{z} \, \hat{\sigma}_{\text{LO}} (Q^2 =
z\tau s) \left\{ - \frac{z}{2} P_{gq} (z) \log \frac{\mu_F^2}{\tau s
    (1-z)^2} \right. \nonumber \\
&& \left. \hspace*{5cm} + \frac{2}{3} z^2 - (1-z)^2
\right\} \label{eq:contrib3} \nonumber\\
\Delta \sigma_{q\bar{q}} &=& \frac{\alpha_s (\mu_R)}{\pi} \int_{\tau_0}^1
d\tau \, \sum_q \frac{d {\cal L}^{q\bar{q}}}{d\tau} 
\int_{\tau_0/\tau}^1 \frac{dz}{z} \, \hat{\sigma}_{\text{LO}} (Q^2 =
z\tau s) \, \frac{32}{27} (1-z)^3 \;. \label{eq:contrib4}
\eeq
Here $s$ denotes the hadronic c.m.~energy and 
\beq
\tau_0 = \frac{4 M_h^2}{s} \;,
\eeq
and the Altarelli-Parisi splitting functions are given by \cite{Altarelli:1977zs},
\beq
P_{gg} (z) &=& 6 \left\{ \left( \frac{1}{1-z} \right)_+ + \frac{1}{z}
  - 2 + z (1-z) \right\} + \frac{33-2N_F}{6} \delta (1-z) \nonumber \\
P_{gq} (z) &=& \frac{4}{3} \frac{1+(1-z)^2}{z} \;,
\eeq
with $N_F=5$ in our case. The factorisation scale of the parton-parton
luminosities $d{\cal L}^{ij}/d\tau$ is denoted by $\mu_F$. The
relative real corrections are not affected by the higher-dimensional
operators. The virtual corrections, however, are changed with respect
to the SM case due to the overall coupling modifications of the top
Yukawa and the trilinear Higgs self-coupling and because of the
additional contributions from the novel effective vertices. The
coefficient $C$ appearing in the virtual corrections is given by 
\bea
C &=& \pi^2 + \frac{33-2N_F}{6} \log \frac{\mu_R^2}{Q^2} \nonumber \\
&+& \mathrm{Re}
\frac{1}{\int_{\hat{t}_-}^{\hat{t}_+} d\hat{t} \left[ |C_\Delta F_1 + F_2|^2 +
    |G_1|^2 + |C_\Delta \tilde{F}_1 + \tilde{F}_2|^2 +|\tilde{G}_{1}|^2 \right]} 
\int_{\hat{t}_-}^{\hat{t}^+} d \hat{t} \Big\{ \nonumber \\ 
&& \left[|C_\Delta F_1 + F_2|^2 + |G_1|^2 \right] \frac{11}{2} +
\left[|C_\Delta \tilde{F}_1 + \tilde{F}_2|^2 + |\tilde{G}_1|^2 \right]
6 \nonumber \\
&& + (C_\Delta F_1 + F_2) \left[-C_\Delta^* \frac{11}{3} c_\Delta +
\frac{11}{3} c_\Box \right] \nonumber \\
&& + (C_\Delta F_1 + F_2) [a_1 (c_t + c_\Delta)^2 + \tilde{a}_1
(\tilde{c}_t + \tilde{c}_\Delta)^2 ] \nonumber \\
&& + (C_\Delta \tilde{F}_1 + \tilde{F}_2) 2 a_2 (c_t + c_\Delta)
(\tilde{c}_t + \tilde{c}_\Delta) \nonumber \\
&& + [a_3 (c_t+c_\Delta)^2 + \tilde{a}_3 (\tilde{c}_t +
\tilde{c}_\Delta)^2 ] \frac{p_T^2}{2\hat{u} \hat{t}} (Q^2 - 2m_h^2)
G_1 \nonumber \\
&& + 2 a_4 (c_t+c_\Delta) (\tilde{c}_t + \tilde{c}_\Delta)
\frac{p_T^2}{2 \hat{u} \hat{t}} (\hat{t}-\hat{u}) \tilde{G}_1
\Big\}\;,
\label{eq:virtualcoeff}
\eea
with
\beq
a_1 = \frac{4}{9} = - a_3 \;, \quad \tilde{a}_1 = -1 = \tilde{a}_3 \;,
\quad a_2 = \frac{2}{3} \;, \quad a_4 = \frac{2}{3} \;, \label{eq:cicoeffs}
\eeq
and the transverse momentum squared
\beq
p_T^2 = \frac{(\hat{t}-M_h^2)(\hat{u}-M_h^2)}{Q^2} - M_h^2 \;.
\eeq
The last four lines in Eq.~(\ref{eq:virtualcoeff}) arise from the
third diagram in Fig.~\ref{fig:nlodiags} (upper), containing the two effective 
Higgs-two-gluon couplings. The remaining terms originate from the
diagrams with gluon loops in Fig.~\ref{fig:nlodiags} (upper).  
In line 3, the factor $11/2$ arises from the matching of the effective theory to
the full theory. This induces the factor $(1+11\alpha_s/(2 \pi))$ in
the CP-even components of the effective couplings of
Fig.~\ref{fig:effcpmixedvertices} for the contributions arising 
from integrating out the top loops,
while the effective couplings not mediated by SM quark loops and the
CP-odd components of the couplings are not affected. The factor 6 
arises in the virtual corrections to Higgs pair production for the
final state projecting on the CP-mixed state, while the purely CP-even and CP-odd
final state projections do not exhibit such factor, {\it cf.}~\cite{Dawson:1998py}. 
Note that we have kept the full top quark mass dependence in the LO amplitude
in the derivation of the coefficient $C$ for the virtual corrections. 

%%%%%%%%%%%%%%%%%%%%%%%%%%%%%%%%%%%%%%%%%%%%%%%%%%%%%%%
\section{Higgs Pair Production in the C2HDM \label{sec:c2dhmcalc}}
In this section we present the NLO QCD corrections to Higgs pair
production in a specific UV complete model. Investigations in
well-defined UV complete models complement the EFT 
approach, as the latter cannot account for NP effects
arising from light resonances. Here we resort to the CP-violating
2-Higgs-Doublet Model, the C2HDM. The extension of the SM Higgs sector
by a complex Higgs doublet naturally fulfills the constraints from the $\rho$
parameter. In the type II 2HDM, furthermore, the two Higgs doublets
couple in the same way to the fermions as in the Minimal
Supersymmetric Extensions of the SM (MSSM). The 2HDM Higgs couplings,
however, are not constrained by supersymmetric relations and thus entail
more substantial deviations from the SM that are still compatible with the
data. In this sense, the 2HDM \cite{Lee:1973iz,Branco:2011iw} is an
important benchmark model for the experimental study of the effects of
extended Higgs sectors. We briefly summarise the basics relevant for
our process and refer to the literature for more details, {\it
  cf.}~{\it e.g.}~\cite{Fontes:2014xva,Muhlleitner:2017dkd}.  

%%%%%%%%%%%%%%%%%%%%%%%%%%%%%%%%%%%%%%%%%%%%%%%%%%%%%%%
\subsection{The C2HDM}
The Higgs potential of a general 2HDM with two $SU(2)_L$ doublets
$\Phi_1$ and $\Phi_2$ and a softly broken discrete $\mathbb{Z}_2$
symmetry reads
\beq
V &=& m_{11}^2 |\Phi_1|^2 + m_{22}^2 |\Phi_2|^2 - m_{12}^2 (\Phi_1^\dagger
\Phi_2 + h.c.) + \frac{\lambda_1}{2} (\Phi_1^\dagger \Phi_1)^2 +
\frac{\lambda_2}{2} (\Phi_2^\dagger \Phi_2)^2 \nonumber \\
&& + \lambda_3
(\Phi_1^\dagger \Phi_1) (\Phi_2^\dagger \Phi_2) + \lambda_4
(\Phi_1^\dagger \Phi_2) (\Phi_2^\dagger \Phi_1) + \frac{\lambda_5}{2}
[(\Phi_1^\dagger \Phi_2)^2 + h.c.] \;. 
\label{eq:c2hdmpot}
\eeq
The absence of tree-level Flavor Changing Neutral Currents (FCNC) is
ensured by the required invariance under the $\mathbb{Z}_2$
transformations $\Phi_1 \to - \Phi_1$ and $\Phi_2 \to \Phi_2$. By
hermiticity all parameters in $V$ are real except for the soft
$\mathbb{Z}_2$ breaking mass parameter $m_{12}^2$ and the quartic
coupling $\lambda_5$. For $\mbox{arg}(m_{12}^2) = \mbox{arg}
(\lambda_5)$, the complex phases of $m_{12}^2$ and $\lambda_5$ can be
absorbed by a basis 
transformation leading to the real or CP-conserving 2HDM, in case the
VEVs of both Higgs doublets are assumed to be
real. Otherwise we are in the CP-violating 2HDM, which depends on
10 real parameters. In the following, we will adopt the conventions of
\cite{Fontes:2014xva} for the C2HDM. The VEVs of
the neutral components of the Higgs doublets developed after EWSB can in
principle be complex if CP violation is allowed. The relative phase
between the VEVs can, however, be rotated away by a global phase
transformation in the field $\Phi_2$ \cite{Ginzburg:2002wt} so that
without loss of generality it can be 
set to zero. After EWSB the two doublets $\Phi_i$ $(i=1,2)$ are
expanded about the real VEVs $v_1$ and $v_2$ and we have
\beq
\Phi_1 = \left(
\begin{array}{c}
\phi_1^+ \\
\frac{v_1 + \rho_1 + i \eta_1}{\sqrt{2}}
\end{array}
\right) \qquad \mbox{and} \qquad
\Phi_2 = \left(
\begin{array}{c}
\phi_2^+ \\
\frac{v_2 + \rho_2 + i \eta_2}{\sqrt{2}}
\end{array}
\right) \;, \label{eq:2hdmdoubletexp}
\eeq
where $\rho_i$ and $\eta_i$ denote the real neutral CP-even and CP-odd
fields, respectively, and $\phi_i^+$ the charged complex
fields. Requiring the minimum of the potential to be located at
\beq
\langle \Phi_i \rangle = \left( \begin{array}{c} 0 \\
    \frac{v_i}{\sqrt{2}} \end{array} \right) \label{eq:2hdmmin}
\eeq
induces the minimum conditions
\beq
m_{11}^2 v_1 + \frac{\lambda_1}{2} v_1^3 + \frac{\lambda_{345}}{2} v_1
v_2^2 &=& m_{12}^2 v_2 \label{eq:cond1} \\
m_{22}^2 v_2 + \frac{\lambda_2}{2} v_2^3 + \frac{\lambda_{345}}{2} v_1^2
v_2 &=& m_{12}^2 v_1 \label{eq:cond2} \\
2\, \mbox{Im} (m_{12}^2) &=& v_1 v_2 \mbox{Im} (\lambda_5)
\;, \label{eq:cond3}
\eeq
where
\beq
\lambda_{345} \equiv \lambda_3 + \lambda_4 + \mbox{Re} (\lambda_5) \;.
\eeq
Equations (\ref{eq:cond1}) and (\ref{eq:cond2}) can be used to trade
the parameters $m_{11}^2$ and $m_{22}^2$ for $v_1$ and
$v_2$, and Eq.~(\ref{eq:cond3}) yields a relation between the two
sources of CP violation in the scalar potential, thus fixing one of
the ten C2HDM parameters. We introduce the mixing angle $\beta$ given
by 
\beq
\tan \beta = \frac{v_2}{v_1} \;,
\eeq
which rotates the two Higgs doublets into the Higgs basis
\cite{Lavoura:1994fv, Botella:1994cs}. 
Defining the CP-odd field $\rho_3 \equiv
-\eta_1 \sin \beta + \eta_2 \cos \beta$ (the orthogonal field
corresponds to the massless Goldstone boson), the neutral mass
eigenstates $H_i$ ($i=1,2,3$) are obtained from the C2HDM basis
$\rho_1$, $\rho_2$ and $\rho_3$ through the rotation
\beq
\left( \begin{array}{c} H_1 \\ H_2 \\ H_3 \end{array} \right) = R
\left( \begin{array}{c} \rho_1 \\ \rho_2 \\ \rho_3 \end{array} \right)
\;.
\label{eq:c2hdmrot}
\eeq
The neutral mass matrix
\beq
({\cal M}^2)_{ij} = \left\langle \frac{\partial^2 V}{\partial \rho_i
  \partial \rho_j} \right\rangle \;,
\label{eq:c2hdmmassmat}
\eeq
is diagonalised by the orthogonal matrix $R$ through
\beq
R {\cal M}^2 R^T = \mbox{diag} (m_{H_1}^2, m_{H_2}^2, m_{H_3}^2) \;.
\eeq
We order the Higgs bosons by ascending mass as $m_{H_1}
\le m_{H_2} \le m_{H_3}$.  Introducing the abbreviations $s_{i} \equiv \sin \alpha_i$ and
$c_{i} \equiv \cos \alpha_i$ with 
\beq
-\frac{\pi}{2} \le \alpha_i < \frac{\pi}{2} \,,
\label{eg:alphas}
\eeq
the mixing matrix $R$ can be parametrised as
\beq
R =\left( \begin{array}{ccc}
c_{1} c_{2} & s_{1} c_{2} & s_{2}\\
-(c_{1} s_{2} s_{3} + s_{1} c_{3})
& c_{1} c_{3} - s_{1} s_{2} s_{3}
& c_{2} s_{3} \\
- c_{1} s_{2} c_{3} + s_{1} s_{3} &
-(c_{1} s_{3} + s_{1} s_{2} c_{3})
& c_{2}  c_{3}
\end{array} \right) \;.
\label{eq:2hdmmatrix}
\eeq
For the 9 independent parameters of the C2HDM we then choose \cite{ElKaffas:2007rq}
\beq
v \approx 246\mbox{ GeV} \;, \quad t_\beta \;, \quad \alpha_{1,2,3}
\;, \quad m_{H_i} \;, \quad m_{H_j} \;, \quad m_{H^\pm} \quad
\mbox{and} \quad \mbox{Re}(m_{12}^2) \;.
\label{eq:indepparams}
\eeq
The $m_{H_i}$ and $m_{H_j}$ denote any of the masses of two among the
three neutral Higgs bosons. The mass of the third Higgs boson is
obtained from the other parameters
\cite{ElKaffas:2007rq}. The analytic relations between the above
parameter set and the coupling parameters $\lambda_i$ of the 2HDM
Higgs potential can be found in \cite{Fontes:2014xva}. For $\alpha_2=
\alpha_3=0$ and $\alpha_1 = \alpha + \pi/2$ the
CP-conserving 2HDM is obtained \cite{Khater:2003wq}. The mass matrix
Eq.~(\ref{eq:c2hdmmassmat}) then becomes block diagonal, $\rho_3$ is
identified with the pure pseudoscalar Higgs boson $A$ and the CP-even
mass eigenstates $h$ and $H$ result from the gauge eigenstates through
the rotation parametrised in terms of the angle $\alpha$, {\it i.e.}
\beq
\left( \begin{array}{c} H \\ h \end{array} \right) =
\left( \begin{array}{cc} c_\alpha & s_\alpha \\ -s_\alpha &
    c_\alpha \end{array} \right) \left( \begin{array}{c} \rho_1 \\
    \rho_2 \end{array} \right) \;.
\eeq

For the computation of the Higgs pair production process we need the
Higgs couplings to two fermions, the $Z$ couplings to two Higgs bosons
and the self-couplings among three Higgs bosons. We allow one type of
fermions to couple only to one Higgs doublet, in order to avoid
tree-level FCNC. This is achieved by imposing a global $\mathbb{Z}_2$
symmetry under which $\Phi_{1,2} \to \mp \Phi_{1,2}$. There are four
phenomenologically different 2HDM types as shown in Table~\ref{tab:yukawatypes}.
\begin{table}
\begin{center}
\begin{tabular}{rccc} \toprule
& $u$-type & $d$-type & leptons \\ \midrule
type I & $\Phi_2$ & $\Phi_2$ & $\Phi_2$ \\
type II & $\Phi_2$ & $\Phi_1$ & $\Phi_1$ \\
lepton-specific & $\Phi_2$ & $\Phi_2$ & $\Phi_1$ \\
flipped & $\Phi_2$ & $\Phi_1$ & $\Phi_2$ \\ \bottomrule
\end{tabular}
\caption{The four different types of Higgs-doublet couplings to
  fermions in the $\mathbb{Z}_2$-symmetric 2HDM. \label{tab:yukawatypes}}
\end{center}
\end{table} 
The Yukawa Lagrangian from which the Higgs couplings to fermions are
derived, reads
\beq
{\cal L}_Y = - \sum_{k=1}^3 \frac{m_f}{v} \bar{\psi}_f \left[ c^e(H_k
  ff) + i c^o(H_k ff) \gamma_5 \right] \psi_f H_k \;, \label{eq:yukawalag}
\eeq
with $\Psi$ denoting the fermion fields of mass $m_f$. The CP-even and
CP-odd Yukawa coupling coefficients $c^e (H_i ff)$ and $c^o (H_i ff)$
were derived in \cite{Fontes:2014xva} and we summarise them in
Table~\ref{tab:yukawacoup}. 
\begin{table}
\begin{center}
\begin{tabular}{rccc} \toprule
& $u$-type & $d$-type & leptons \\ \midrule
type I & $\frac{R_{k2}}{s_\beta} - i \frac{R_{k3}}{t_\beta} \gamma_5$
& $\frac{R_{k2}}{s_\beta} + i \frac{R_{k3}}{t_\beta} \gamma_5$ &
$\frac{R_{k2}}{s_\beta} + i \frac{R_{k3}}{t_\beta} \gamma_5$ \\
type II & $\frac{R_{k2}}{s_\beta} - i \frac{R_{k3}}{t_\beta} \gamma_5$
& $\frac{R_{k1}}{c_\beta} - i t_\beta R_{k3} \gamma_5$ &
$\frac{R_{k1}}{c_\beta} - i t_\beta R_{k3} \gamma_5$ \\
lepton-specific & $\frac{R_{k2}}{s_\beta} - i \frac{R_{k3}}{t_\beta} \gamma_5$
& $\frac{R_{k2}}{s_\beta} + i \frac{R_{k3}}{t_\beta} \gamma_5$ &
$\frac{R_{k1}}{c_\beta} - i t_\beta R_{k3} \gamma_5$ \\
flipped & $\frac{R_{k2}}{s_\beta} - i \frac{R_{k3}}{t_\beta} \gamma_5$
& $\frac{R_{k1}}{c_\beta} - i t_\beta R_{k3} \gamma_5$ &
$\frac{R_{k2}}{s_\beta} + i \frac{R_{k3}}{t_\beta} \gamma_5$ \\ \bottomrule
\end{tabular}
\caption{The Yukawa coupling coefficients of the C2HDM Higgs
  bosons $H_k$ corresponding to the expression 
  $[c^e(H_k ff) +i c^o (H_k ff) \gamma_5]$ in
  Eq.~(\ref{eq:yukawalag}). \label{tab:yukawacoup}}
\end{center}
\end{table}
In the CP-conserving 2HDM the $Z$ boson would only couple to the
CP-mixed Higgs pair combination $hA$ or $HA$. In the case of CP violation it can
couple to any pair of Higgs bosons $H_i H_j$. In terms of the
$SU(2)_L$ gauge coupling $g$ and the cosine of the Weinberg angle
$\theta_W$ the Feynman rule for the coupling $Z^\mu H_i H_j$ is given by
\beq
-\frac{g}{2 \cos \theta_W} (p_{H_i} - p_{H_j})^\mu \, c(ZH_i H_j) \;,
\eeq
where the four-momenta of both Higgs bosons, $p_{H_{i,j}}$, are taken
as incoming. The coupling coefficients $c(ZH_i H_j)$, parametrised by the mixing
matrix elements $R_{ij}$ and the ratio of the two VEVs, $\tan\beta$, read
\beq
c(ZH_i H_j) = (R_{j2} \cos\beta - R_{j1} \sin\beta) R_{i3} +
(-R_{i2} \cos\beta + R_{i1} \sin\beta) R_{j3} \;.
\eeq
Note, that the coupling coefficient $c(ZH_i H_j)$ becomes zero for $i=j$. 
The trilinear Higgs self-couplings $\lambda_{H_i H_j H_k}$ are quite
lengthy and we will not list them here explicitly. In the
CP-conserving limit for a SM-like Higgs boson $h$, the trilinear
self-coupling approaches $3M_h^2/M_Z^2$. 

%%%%%%%%%%%%%%%%%%%%%%%%%%%%%%%%%%%%%%%%%%%%%%%%%%%%%%%
\subsection{The NLO QCD Corrections in the C2HDM}
\begin{figure}[b]
\includegraphics[width=\textwidth]{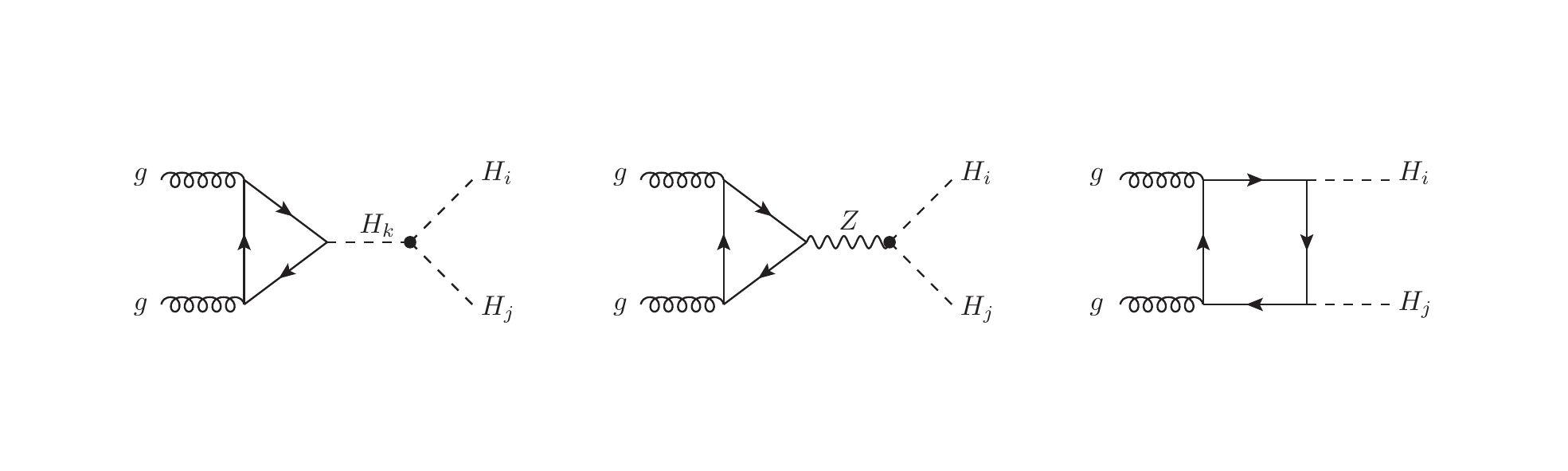}
%\begin{minipage}{0.33\textwidth}
% \includegraphics[width=\textwidth]{fig2-1}
%\end{minipage}
%\begin{minipage}{0.33\textwidth}
% \includegraphics[width=\textwidth]{fig2-1}
%\end{minipage}
%\begin{minipage}{0.33\textwidth}
% \includegraphics[width=\textwidth]{fig2-3}
%\end{minipage}
%
\vspace*{-1.5cm}
\caption{Generic diagrams contributing to C2HDM Higgs pair production in
  gluon fusion at LO. \label{fig:loc2hdmdiags}} 
\end{figure}

The diagrams contributing to the LO production of a C2HDM Higgs pair
$H_i H_j$ are depicted in Fig.~\ref{fig:loc2hdmdiags}. In contrast to the EFT approach,
the cross section does not receive
contributions from the effective couplings, obtained from integrating
out heavy states. Furthermore, as we have now three
CP-violating Higgs states $H_i$, we can have different combinations of
Higgs pairs in the final state, and in the first diagram of
Fig.~\ref{fig:loc2hdmdiags} we have to sum over all three possible Higgs boson
exchanges $H_k$ ($k=1,2,3$). Finally, we have an additional diagram contributing to
Higgs pair production where a virtual $Z$ boson couples to the
triangle and subsequently decays into a Higgs pair, {\it cf.}~second
diagram in Fig.~\ref{fig:loc2hdmdiags}. This diagram does not
contribute for equal Higgs bosons in the final state, as the coupling
coefficient $c(ZH_iH_j)$ vanishes in this case. 
The LO partonic cross section for the
production of the Higgs pair $H_iH_j$ ($i,j=1,2,3$) can then be cast into the form
\begin{eqnarray}
\hat{\sigma}_{\text{LO}} (gg \to H_i H_j) &=& \int_{\hat{t}_-}^{\hat{t}_+} 
d\hat{t} \, \frac{G_F^2 \alpha_s^2(\mu_R)}{256 (2\pi)^3 (1+\delta_{ij})} \,
\left[
\left| \left( \sum_{k=1}^3 C_{\Delta,ij}^k F_1^k \right) + F_{2,ij}
\right|^2   + |G_{1,ij}|^2  \right. 
\nonumber \\
&+& \left.  \left| \left( \sum_{k=1}^3 
  C_{\Delta,ij}^k \tilde{F}_{1}^k \right) + C_{\Delta,ij}^Z
\tilde{F}_1^Z + \tilde{F}_{2,ij} \right|^2 +
|\tilde{G}_{1,ij}|^2
 \right] \;, \label{eq:loc2hdmsigma} 
\end{eqnarray}
where
\begin{eqnarray}
C_{\Delta,ij}^k &=& \lambda_{H_k H_i H_j} \frac{M_Z^2}{Q^2-M_{H_k}^2 + i
  M_{H_k} \Gamma_{H_k}} \nonumber \\
C_{\Delta,ij}^Z &=& -c(ZH_iH_j) \frac{M_Z^2}{Q^2-M_Z^2+ i M_Z
                    \Gamma_Z}
\label{eq:triangles}
\end{eqnarray}
and
\begin{eqnarray}
F_1^k &=& c^e_k F^e_\Delta \label{eq:formfac1} \nonumber \\
\tilde{F}_1^k &=& c^o_k F^o_\Delta \nonumber \\
\tilde{F}_1 ^Z &=& a_t F^Z_\Delta \nonumber \\
F_{2,ij} &=& c^e_i c^e_j F_\Box^e +
c^o_i c^o_j F_\Box^o \nonumber \\
\tilde{F}_{2,ij} &=& (c^e_i c^o_j +c^e_j
                     c^o_i) F_\Box^{m} \nonumber \\
G_{1,ij} &=& c^e_i c^e_j G_\Box^e +
c^o_i c^o_j G_\Box^o \nonumber \\
\tilde{G}_{1,ij} &=&  (c^e_i c^o_j +c^e_j
                     c^o_i) G_\Box^{m} \label{eq:formfac7}
\end{eqnarray}
where $a_t =1$ denotes the axial charge of the top quark in the loop, 
and where we have used the short-hand notation
\beq
c^e_i \equiv c^e (H_i tt) \qquad \mbox{and} \qquad
c^o_i \equiv c^o (H_i tt)  \;.
\eeq
In Eq.~(\ref{eq:triangles}) the $\Gamma_{H_k}$ and $\Gamma_Z$ denote
the total widths of the Higgs boson $H_k$ and the $Z$ boson,
respectively.\footnote{Higgs pair production at LO in the
  C2HDM has been investigated in Ref.~\cite{Bian:2016awe}, but only
  for SM-like Higgs pairs, {\it i.e.}~equal final states, where the
  $Z$ exchange in the $s$-channel diagram 
  does not contribute.} For a given set of input parameters we obtain the total
Higgs width with a private version of {\tt HDECAY}
\cite{Djouadi:1997yw,Djouadi:2006bz}, adapted to the C2HDM, that will 
be published in a forthcoming paper. The form factors appearing in
Eqs.~(\ref{eq:formfac7}) are the same as the ones
given in section~\ref{sec:eftnlocalc}, apart from $F_\Delta^Z$. 
They can all be found in the appendix of Ref.~\cite{Plehn:1996wb}. \s

In the computation of the NLO QCD corrections in the heavy quark limit
we again consistently neglect the bottom quark loops in the
following, and we use the Feynman rules for the effective couplings
between one and two Higgs bosons to two gluons in the heavy top quark limit. 
\begin{figure}[h]
\begin{minipage}{.25\textwidth}
\includegraphics{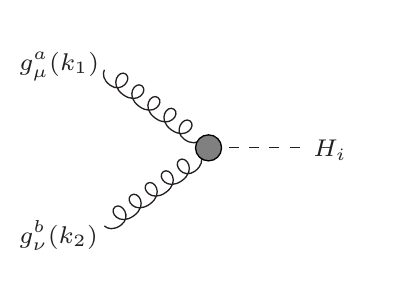}
\end{minipage}
\begin{minipage}{.75\textwidth}
\begin{flushleft}
\beq
i \delta^{ab} \frac{\alpha_s}{\pi v} \Big\{ && \hspace*{-0.6cm}
\frac{1}{3}[k_1^\nu k_2^\mu - (k_1 \cdot k_2) g^{\mu\nu}] \left[ c^e_i
  \left(1+\frac{11}{4} \frac{\alpha_s}{\pi} \right) \right] \nonumber
\\ - && \hspace*{-0.6cm} \frac{1}{2}
\epsilon^{\mu\nu\sigma\rho}k_{1\sigma}k_{2\rho} \, c^o_i \Big\} 
\nonumber
\eeq
  \end{flushleft}
\end{minipage}\\
\begin{minipage}{.25\textwidth}
\includegraphics{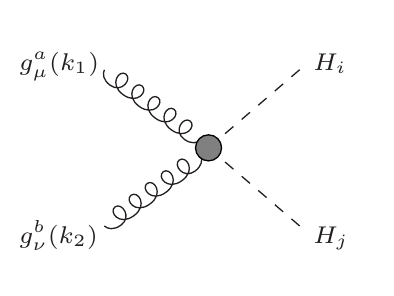}
\end{minipage}
\begin{minipage}{.75\textwidth}
\begin{flushleft}
\beq
i \delta^{ab} \frac{\alpha_s}{\pi v^2} \Big\{ && \hspace*{-0.6cm} \frac{1}{3}[k_1^\nu k_2^\mu - (k_1 \cdot
k_2) g^{\mu\nu}] \left[ (-c^e_i c^e_j +c^o_i c^o_j) \left(1+\frac{11}{4}
    \frac{\alpha_s}{\pi} \right) \right] \nonumber \\ 
+ && \hspace*{-0.6cm} \frac{1}{2} 
\epsilon^{\mu\nu\sigma\rho}k_{1\sigma}k_{2\rho}
\left[ c^e_i c^o_j + c^e_j c^o_i \right] \Big\} 
\nonumber
\eeq \\
\end{flushleft}
\end{minipage}
\caption{Feynman rules for the effective two-gluon
  couplings to one (upper) 
  and two (lower) CP-violating Higgs bosons in the heavy quark limit,
  including NLO QCD corrections. The four-momenta of the
  gluons, $k_1$ and $k_2$, are taken as both incoming or both
  outgoing. \label{fig:effrulesc2hdm}} 
\end{figure}
They are given in Fig.~\ref{fig:effrulesc2hdm}. For $H_i=H_j\equiv h$,
these rules can be obtained from the corresponding ones in the EFT approach,
Fig.~\ref{fig:effcpmixedvertices}, by making the replacements
\beq
c_t \to c^e_i \;,\; \tilde{c}_t \to c^o_i \;,\; 
\{ c_g \;,\; c_{gg} \;,\; c_{tt} \;,\; \tilde{c}_g \;,\;
\tilde{c}_{gg} \;,\; \tilde{c}_{tt}  \} \to 0 \;.
\eeq
%Note, that the factor $(1+\delta_{ij})$ accounts for the symmetry
%factor in case of two identical Higgs bosons coupling to the two
%gluons. 
Analogously to the NLO corrections in the EFT, the NLO
corrections can be cast into the form of Eq.~(\ref{eq:nlocxn}). The individual
contributions are given as in Eqs.~(\ref{eq:contrib1})
with the LO cross section replaced by the 2HDM result in 
Eq.~(\ref{eq:loc2hdmsigma}) and the factor $C$ for the virtual
corrections given by
\bea
C && = \pi^2 + \frac{33-2N_F}{6} \log \frac{\mu_R^2}{Q^2} \label{eq:nloc2hdm} \\
&& + \mathrm{Re}
\frac{1}{\int_{\hat{t}_-}^{\hat{t}_+} d\hat{t} \left[
\left| C_{\Delta,ij}^k F_1^k + F_{2,ij}
\right|^2   + |G_{1,ij}|^2 + \left| 
  C_{\Delta,ij}^k \tilde{F}_{1}^k + C_{\Delta,ij}^Z
\tilde{F}_1^Z + \tilde{F}_{2,ij} \right|^2 +
|\tilde{G}_{1,ij}|^2  \right]} \times \nonumber \\
&&\int_{\hat{t}_-}^{\hat{t}^+} d \hat{t} \Big\{ 
\left[ \left| C_{\Delta,ij}^k F_1^k + F_{2,ij}
\right|^2 + |G_{1,ij}|^2 \right] \frac{11}{2} +
\left[ \left| 
  C_{\Delta,ij}^k \tilde{F}_{1}^k + C_{\Delta,ij}^Z
\tilde{F}_1^Z + \tilde{F}_{2,ij} \right|^2 + |\tilde{G}_{1,ij}|^2\right]
6 \nonumber \\
&& + (C_{\Delta,ij}^k F_1^k + F_{2,ij}) [a_1 (c^e_i c^e_j) + \tilde{a}_1
(c^o_i c^o_j)] + (C_{\Delta,ij}^k \tilde{F}_1^k + C_{\Delta,ij}^Z
\tilde{F}_1^Z+ \tilde{F}_{2,ij}) 
a_2 (c^e_i c^o_j +c^e_j c^o_i)
\nonumber \\
&& + [a_3 (c^e_i c^e_j) + \tilde{a}_3
(c^o_i c^o_j)] \frac{p_T^2}{2\hat{u} \hat{t}} (Q^2 - M_{H_i}^2 - M_{H_j}^2) 
G_{1,ij} 
%\nonumber \\
+ a_4 (c^e_i c^o_j +c^e_j c^o_i)
\frac{p_T^2}{2 \hat{u} \hat{t}} (\hat{t}-\hat{u}) \tilde{G}_{1,ij}
\Big\}\;.
\nonumber
\eea
In Eq.~(\ref{eq:nloc2hdm}) we have implicitly assumed summation over
same indices. The factors $a_i$ and $\tilde{a}_i$ are given in
Eq.~(\ref{eq:cicoeffs}) and 
\beq
p_T^2 = \frac{(\hat{t}-M_{H_i}^2)(\hat{u}-M_{H_i}^2)}{Q^2} - M_{H_i}^2 \;.
\eeq

%%%%%%%%%%%%%%%%%%%%%%%%%%%%%%%%%%%%%%%%%%%%%%%%%%%%%%%
\section{Numerical Analysis \label{sec:numerical}}
We have implemented the LO and NLO Higgs pair production cross
sections both for the EFT approach including CP violation and for the
C2HDM in the Fortran program {\tt HPAIR} \cite{hpair}. For our
numerical analysis we have chosen the c.m.~energy
$\sqrt{s}=14$~TeV.
% and for comparison also the very high energy  
%option $\sqrt{s}=100$~TeV. 
The Higgs boson mass has been set equal to
$M_h=125$~GeV \cite{Aad:2015zhl} and the top quark
mass has been chosen as $m_t=173.2$~GeV. We have adopted the CT14 parton
densities \cite{Dulat:2015mca} for the LO and NLO cross sections with $\alpha_s 
(M_Z)=0.118$ at LO and NLO. The renormalisation scale has been set
equal to $M_{HH}/2$, where $M_{HH}$ generically denotes the invariant mass of the
final state Higgs pair. Consistent with the application of the
heavy top quark limit in 
the NLO QCD corrections we neglect the bottom quark loops in the LO
cross section. \s

%%%%%%%%%%%%%%%%%%%%%%%%%%%%%%%%%%%%%%%%%%%%%%%%%%%%%%%
\subsection{Impact of CP Violation on NLO QCD Higgs Pair Production in
  the EFT Approach \label{sec:eftnloqcdanalysis}} 
\begin{figure}[t!]
\begin{center}
\includegraphics[scale=0.9]{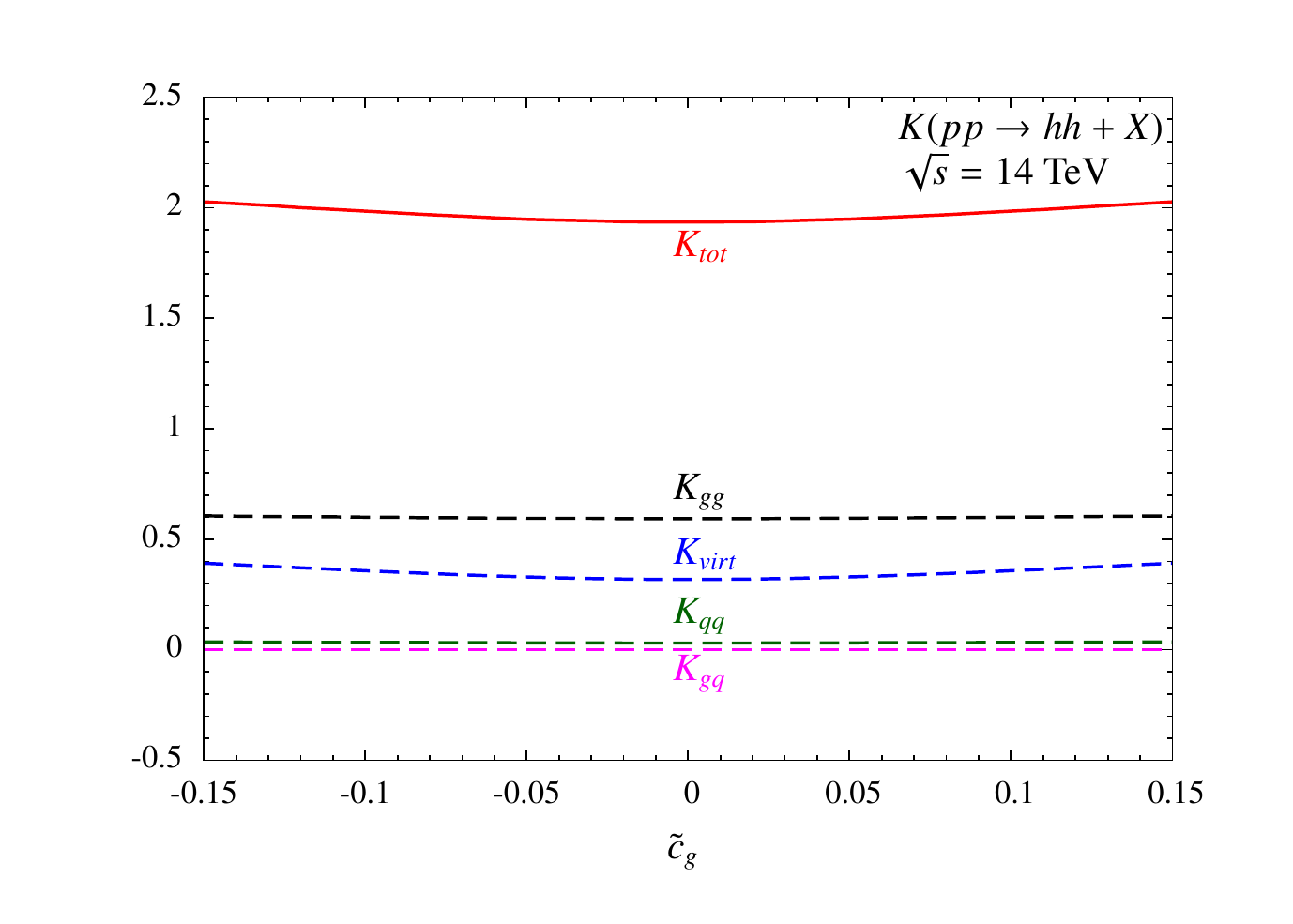} \\[0cm]
\includegraphics[scale=0.9]{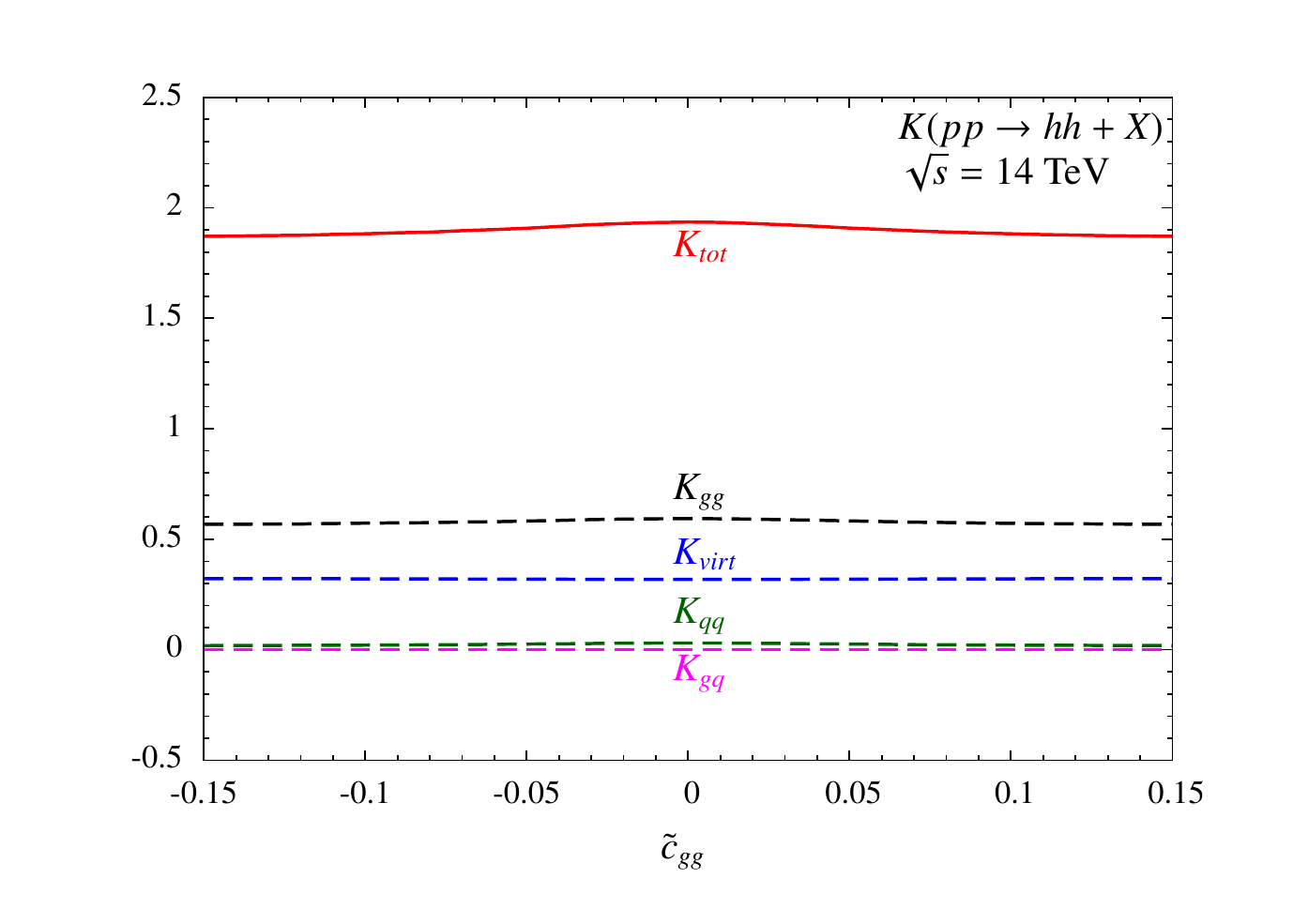}
%\vspace*{-0.5cm}
  \caption{$K$-factors of the QCD-corrected gluon fusion cross
    section $\sigma (pp \to hh +X)$ at the LHC with c.m. energy
    $\sqrt{s}=14$~TeV.  The dashed lines correspond to the individual
    contributions of the QCD corrections given in 
    Eq.~(\ref{eq:nlocxn}), {\it
      i.e.}~$K_i=\Delta \sigma_i/\sigma_{\text{LO}}$
    ($i=\text{virt},gg, gq, q\bar{q}$). Upper: variation of 
    $\tilde{c}_g$ and $\tilde{c}_{gg} =0$; lower: 
    variation of $\tilde{c}_{gg}$ and $\tilde{c}_{g}=0$. The remaining
    couplings have been set to their SM values. \label{fig:varcgandcgg}} 
\end{center}
\vspace*{-0.5cm}
\end{figure}

\begin{figure}[ht!]
\begin{center}
\includegraphics[scale=0.8]{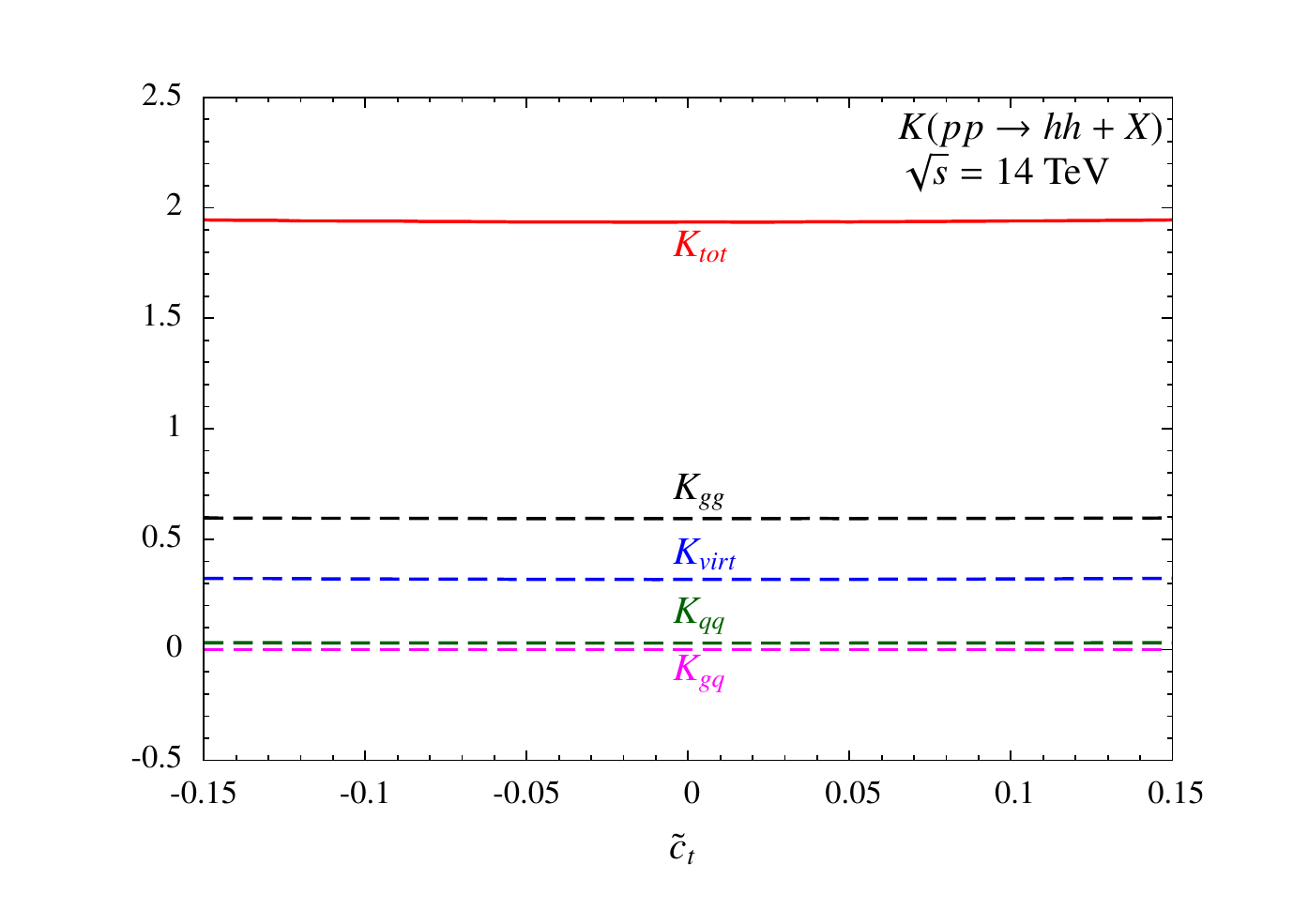}
%\vspace*{-1.3cm}
 \caption{Same as Fig.~\ref{fig:varcgandcgg}, but here
    $\tilde{c}_{t}$ is varied, while the remaining
    values are set to their SM values. 
\label{fig:varct}}
\end{center}
\end{figure}

\begin{figure}[t!]
\begin{center}
\includegraphics[scale=0.8]{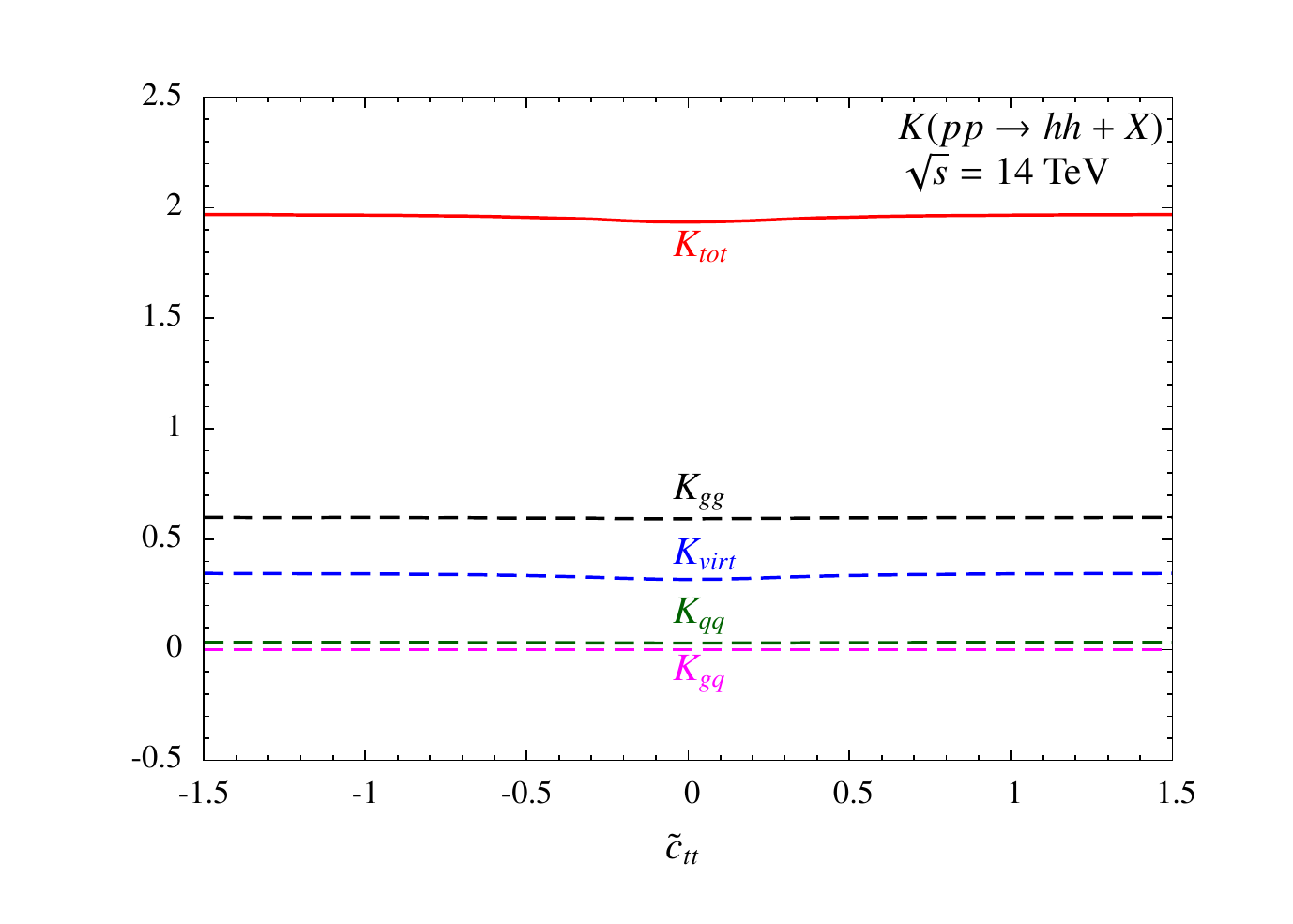}
%\vspace*{-1.3cm}
  \caption{Same as Fig.~\ref{fig:varcgandcgg}, but here
    $\tilde{c}_{tt}$ is varied, while the remaining
    values are set to their SM values. 
\label{fig:varctt}}
\end{center}
\end{figure}

The impact of the new CP-violating couplings in the EFT approach 
on the QCD corrections can be read off
Figs.~\ref{fig:varcgandcgg}-\ref{fig:varctt}.\footnote{The impact of 
  the effects from dimension-6 operators in the CP-conserving case has
been studied in \cite{Grober:2015cwa}.}  
They display the $K$-factor, which is defined as the ratio of the NLO
and LO cross sections, $K=\sigma_{\text{NLO}}/\sigma_{\text{LO}}$,
where the parton densities and the strong couplings $\alpha_s$ are
taken at NLO and LO, respectively.
%\footnote{\textcolor{red}{Note that
%    the individual $K$-factors not add up to the total $K$-factor as
%    they have all been evaluated with NLO parton
%  densities and $\alpha_s$. {\bf True?}}} 
Deviations from the SM $K$-factor
arise both in the virtual and the real corrections. In the virtual
corrections they emerge from the terms in the curly brackets of the
coefficient $C$, Eq.~(\ref{eq:virtualcoeff}). In the real corrections
the different weights in the $\tau$ integration due to the
modified LO cross section induce deviations from the SM. 
In Fig.~\ref{fig:varcgandcgg} (upper) all couplings are set to their SM
values, except for $\tilde{c}_g$. The CP-violating component of the
new contact interaction of 
the Higgs boson to two gluons is varied in the range $-0.15 \le
\tilde{c}_g \le 0.15$. The chosen rather large range is due to
illustrative purposes. In the lower plot, we instead set $\tilde{c}_{g}$ to zero and
vary $\tilde{c}_{gg}$ in the range $-0.15 \le \tilde{c}_{gg} \le 0.15$
while the remaining values are chosen as in the SM. 
The upper plot shows that the CP-violating new interaction $\tilde{c}_g$
induces a variation of the $K$-factor between the SM-value
1.94 and 2.03 in the chosen range\footnote{Note that we find a slightly
  higher SM $K$-factor than in \cite{Grober:2015cwa} where
  $K_{\text{tot}}=1.89$. This is due to the different
  renormalisation scale ($\mu_R =M_{HH}$) and a different
pdf-set (MSTW08) used there. In \cite{Grober:2016wmf}, where we used
the same renormalisation scale as here, but another pdf set (MSTW08), we
found the SM $K$-factor $K_{\text{tot}}=1.71$.}. We define the maximal deviation of
the $K$-factor away from the SM value $K^{\text{SM}}=1.94$, induced by the
coupling $c_x$, as  
\beq
\delta_{\,\text{max}}^{\,K,c_{x}} =
\frac{\mbox{max}|K^{c_{x}}-K^{\text{SM}}|}{K^{\text{SM}}} \;.
\eeq 
The impact on the total cross section is measured by the quantity
\beq
\delta_{\,\text{max}}^{\,\sigma,c_{x}} =
\frac{\mbox{max}|\sigma^{c_{x}}-\sigma^{\text{SM}}|}{\sigma^{\text{SM}}} \;.
\eeq
With these definitions, we find for $\tilde{c}_g = \pm 0.15$,
\beq
\delta_{\,\text{max}}^{\,K,\tilde{c}_{g}} = 0.048 \;.
\eeq
While the effect on the $K$-factor is small, the impact on the
total cross section is significantly more important, where we have for
$\tilde{c}_g= \pm 0.15$ 
\beq
\delta_{\,\text{max}}^{\,\sigma,\tilde{c}_{g}} = 0.519 \;.
\eeq
%
%When also $c_g$ is set to a non-zero positive value with $c_g=0.15$, we find that the
%effects of $\tilde{c}_g$ and $c_g$ add on each other, and both
%$\delta_{\text{max}}^{K,c_g}$ and $\delta_{\text{max}}^{\sigma,c_g}$
%are increased. 
A non-zero $\tilde{c}_{gg}$ has a smaller effect on the $K$-factor and
induces for $\tilde{c}_{gg}=\pm 0.15$
\beq
\delta_{\,\text{max}}^{\,K,\tilde{c}_{gg}} = 0.033 \;.
\eeq
in the investigated range of variation. The impact on the cross
section on the other hand is much more important, with 
\beq
\delta_{\,\text{max}}^{\,\sigma,\tilde{c}_{gg}} = 4.41 \;.
\eeq
If in addition the CP-even contact interactions $c_g$
and/or $c_{gg}$ are set to non-zero values, this can lead to smaller or
larger values of $\delta_{\text{max}}^{K}$ and
$\delta_{\text{max}}^{\sigma}$, depending on the chosen values. \s

For the variation of $\tilde{c}_t$ in the range $-0.15 \le \tilde{c}_t
\le 0.15$ with all other couplings set to their SM values, {\it
  cf.}~Fig.~\ref{fig:varct}, we find for $\tilde{c}_t=\pm 0.15$ 
\beq
\delta_{\,\text{max}}^{\,K,\tilde{c}_{t}} = 0.005 \;.
\eeq
and
\beq
\delta_{\,\text{max}}^{\,\sigma,\tilde{c}_{t}} = 0.198 \;.
\eeq
Both the impact on the $K$-factor and the total cross section is
small. The effect of $\tilde{c}_t$ in the numerator of $C$,
Eq.~(\ref{eq:virtualcoeff}), almost cancels against the one in the
denominator. \s

The impact of the variation of $\tilde{c}_{tt}$ finally, is shown in
Fig.~\ref{fig:varctt}. It is of the per-cent order on the $K$-factor,
with
\beq
\delta_{\,\text{max}}^{\,K,\tilde{c}_{tt}} = 0.018 
\eeq
for $\tilde{c}_{tt}=\pm 1.5$. With
\beq
\delta_{\,\text{max}}^{\,\sigma,\tilde{c}_{tt}} = 14.23 
\eeq
the change of the cross section is substantial. For
$\tilde{c}_{tt}=\pm 0.2$, however, it
induces with $\delta^{\,\sigma,\tilde{c}_{tt}} = 0.25$ similar changes
as the other CP-violating couplings. \s

In summary, the new CP-violating couplings change the $K$-factor by a
few per cent only, while the total cross section itself is affected
much more significantly. We have varied here, however, the new
couplings only one by one away from the SM values. The combined effect
of all dimension-6 couplings, both CP-even and CP-odd, might induce
more substantial deviations in the $K$-factor.

%%%%%%%%%%%%%%%%%%%%%%%%%%%%%%%%%%%%%%%%%%%%%%%%%%%%%%%
\subsection{Impact of CP Violation in the 2HDM on NLO QCD Higgs pair
  production} 
After applying the minimisation conditions of the Higgs potential, we
are left with 9 input parameters for the C2HDM, which we choose as
given in Eq.~(\ref{eq:indepparams}). For our numerical analysis we
adopt as starting point a scenario that is compatible with all 
the relevant constraints. It is obtained from the sample generated in
Ref.~\cite{Muhlleitner:2017dkd}, where we used the tool {\tt ScannerS}
\cite{Coimbra:2013qq,ScannerS} to perform a scan over the input
parameters and check for the 
experimental and theoretical constraints: The potential has been 
required to be bounded from below, the EW vacuum has been ensured to be
a global minimum and it has been checked that tree-level perturbative
unitarity holds. The relevant flavour constraints have been applied
and agreement with the EW precision observables has been
verified. The mass of one of the neutral Higgs bosons has been set to
125~GeV, and compatibility with the Higgs exclusion bounds for the
non-SM Higgs bosons and the individual signal strength fits for the
125~GeV Higgs boson has been checked. Finally, the constraint arising
from the measurement of the electron EDM, which is the most
constraining of the EDMs, has been taken into account. 
For the numerical analysis applied here, we resort to the C2HDM type
II. While this choice does not affect the couplings involved in Higgs
pair production, as the bottom loop contribution has 
consistently been neglected, this is 
relevant for the parameter ranges that are still allowed after
applying the constraints. We included the latest bound on the
charged Higgs boson mass, $m_{H^\pm} > 580$~GeV, of
Ref.~\cite{Misiak:2017bgg}, resulting from the recently updated
analysis by the Belle collaboration  of the inclusive weak radiative $B$-meson 
decays \cite{Belle:2016ufb}. For
further details on the scan and the applied constraints, we refer to
Ref.~\cite{Muhlleitner:2017dkd}. The chosen scenario, finally, is
given by
\beq
&& \alpha_1 = 0.853 \;, \; \alpha_2 = -0.103 \;,\; \alpha_3 = 0.0072
\;, \; \tan\beta = 0.969 \;, \; \mbox{Re}(m_{12}^2) = 70957 \mbox{ GeV}^2
\;,\nonumber \\ 
&& m_{H_1} = 125 \mbox{ GeV} \;,\; m_{H_2} = 377.6 \mbox{ GeV} \;,\;
m_{H^\pm} = 709.7 \mbox{ GeV} \;, \label{eq:scenario}
\eeq 
and the EW VEV $v$ is obtained from the Fermi constant. 
Note, that due to the small value of $\tan\beta$, the omitted bottom
loop contribution is negligible. This scenario leads to the $H_3$ mass
\beq
m_{H_3} = 711.5 \mbox{ GeV}\;,
\eeq
and the CP-even and CP-odd coupling coefficients of the SM-like Higgs
boson $H_1$ to the top quarks $c^e (H_1 t\bar{t})=1.077$ and $c^o(H_1
t\bar{t})= -0.106$. Defining the pseudoscalar admixture to the Higgs boson $H_i$ by
\beq
\Psi_i \equiv (R_{i3})^2 \;,
\eeq
we find a pseudoscalar admixture to $H_1$ of $\Psi_i = 1.06\%$. In
\cite{Muhlleitner:2017dkd} it was shown that pseudoscalar admixtures to
the SM-like Higgs boson of up to 10\% are still compatible with all
constraints. The total widths of the neutral Higgs bosons are obtained as
\beq
\Gamma^{\text{tot}}_{H_1} = 3.695 \cdot 10^{-3} \mbox{ GeV}\;, \quad
\Gamma^{\text{tot}}_{H_2} = 2.664 \mbox{ GeV}\quad \mbox{and} \quad
\Gamma^{\text{tot}}_{H_3} = 75.66 \mbox{ GeV}\;.
\eeq
Starting from this scenario, we vary $\alpha_2$, one of the two angles inducing CP
violation, in the range
\beq
-0.13 \le \alpha_2 \le 0.15 \;.
\eeq 
The lower limit is given by the fact that $m_{H_3}^2$ becomes
negative below this value. This $\alpha_2$ range corresponds to a change in
the SM-like Higgs boson CP-even and CP-odd coupling coefficients to the top
quarks in the ranges
\beq
1.073 \le c^e (H_1 t\bar{t}) \le 1.082 \quad \mbox{and} \quad
-0.134 \le c^o (H_1 t\bar{t}) \le 0.154 \;,
\eeq 
where the maximum value of $c^e (H_1 t\bar{t})$ is obtained for
$\alpha_2=0$. The pseudoscalar admixture varies in the range $0\%
\le \Psi_1 \le 2.23\%$. 
% 1.68\% for -0.13
Note that the scenarios obtained in this way are not necessarily compatible
with all applied constraints. Also the total width of $H_3$ becomes
very large for $\alpha_2 \le -0.11$. As we want to investigate the 
impact of CP violation on NLO QCD Higgs pair production, we still
allow these scenarios for illustrative purposes.
%From these
%input parameters the mixing matrix elements $R_{ij}$,
%Eq.~(\ref{eq:2hdmmatrix}), are calculated 
%internally in {\tt hpair.f} and subsequently all needed C2HDM Higgs
%couplings. 
With these coupling coefficients we are near the SM case for the
CP-even component of the top Yukawa coupling and the variation of
$\tilde{c}_t$ is comparable to the one in 
the EFT approach, where we chose $-0.15 \le \tilde{c}_t \le 0.15$ in
Fig.~\ref{fig:varct}. 
Contrary to the EFT approach, however, in the C2HDM we have additional
Higgs bosons. These and the $Z$ boson contribute in the triangle
diagrams of Higgs pair production, {\it cf.}~diagrams 1 and 2 in
Fig.~\ref{fig:loc2hdmdiags}. Depending on the investigated final
state, the masses of the heavy neutral Higgs bosons, $m_{H_2}$ and/or 
$m_{H_3}$, may be large enough that $H_2$ and/or $H_3$ decay on-shell
into the final state Higgs pair. This can induce resonantly enhanced
Higgs pair production cross sections for the $H_1 H_1$, $H_1 H_2$ or
$H_2 H_2$ final states, provided the branching ratio of the resonantly
produced Higgs boson into the Higgs pair final state is large enough. \s

\begin{figure}[t!]
\begin{center}
\includegraphics[scale=0.8]{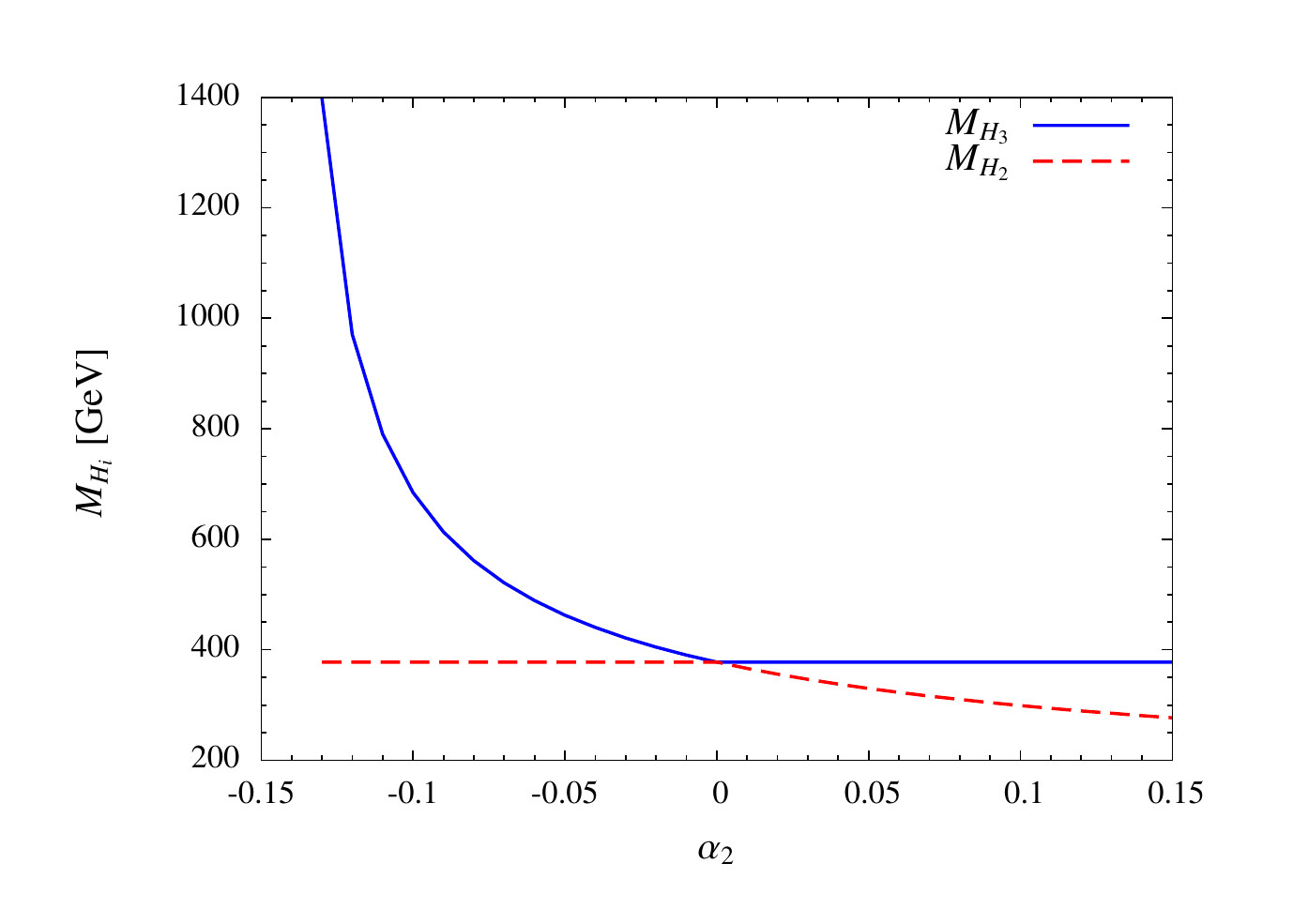}
%\vspace*{-1.3cm}
  \caption{The masses of $H_2$ (red/dashed) and $H_3$ (blue/full) as a
    function of $\alpha_2$.  
\label{fig:h2h3masses}}
\end{center}
\end{figure}
\begin{figure}[t!]
\begin{center}
\includegraphics[scale=0.8]{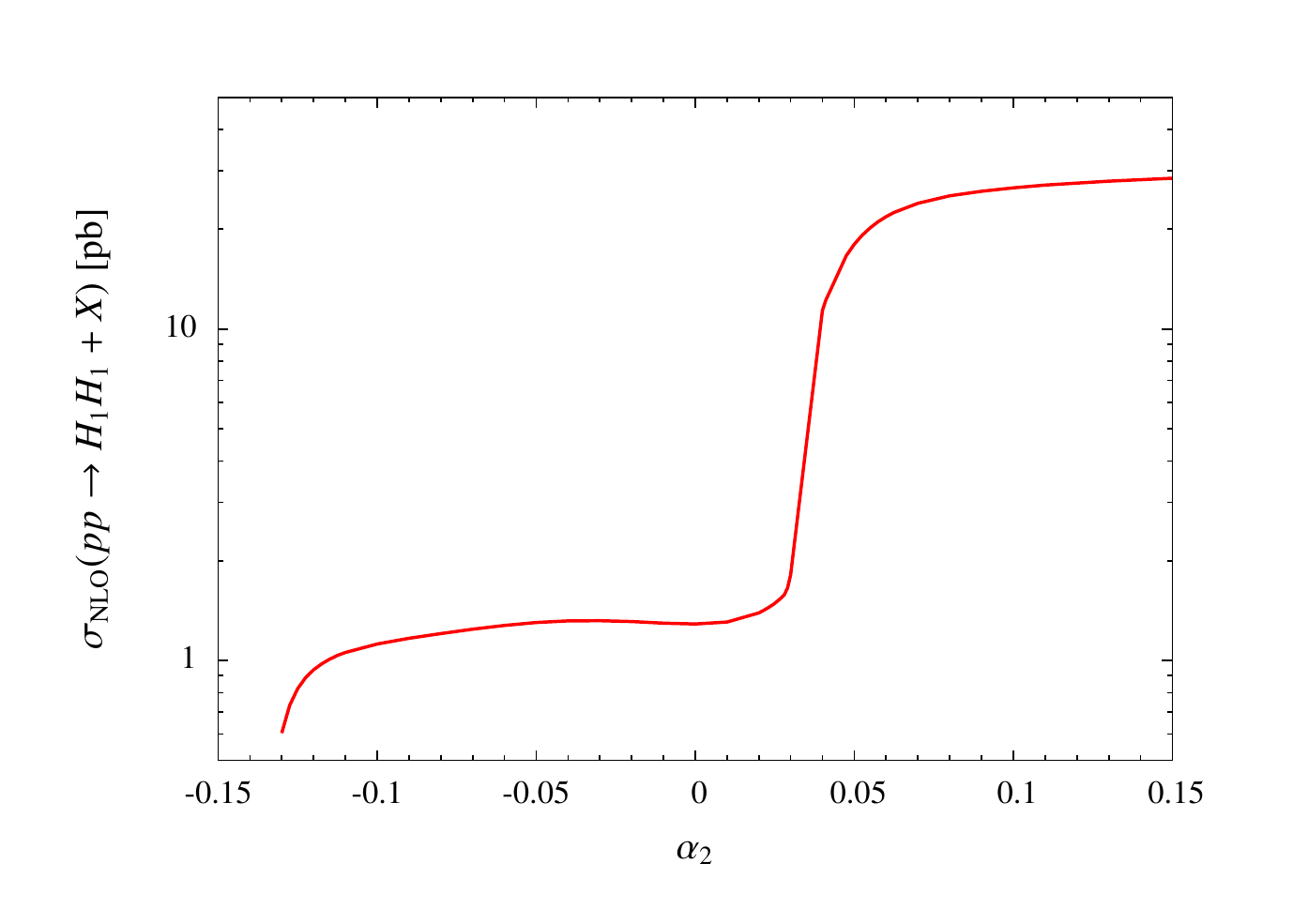}
%\vspace*{-1.3cm}
  \caption{The cross section $\sigma(pp \to H_1 H_1 + X)$ at NLO QCD
    at the LHC with c.m.~energy $\sqrt{s} = 14$~TeV as a function of
    the C2HDM mixing angle $\alpha_2$. All other parameters are given
    as in Eq.~(\ref{eq:scenario}).
\label{fig:nloc2hdmcxn}}
\end{center}
\end{figure}
\begin{figure}[t!]
\begin{center}
\includegraphics[scale=0.8]{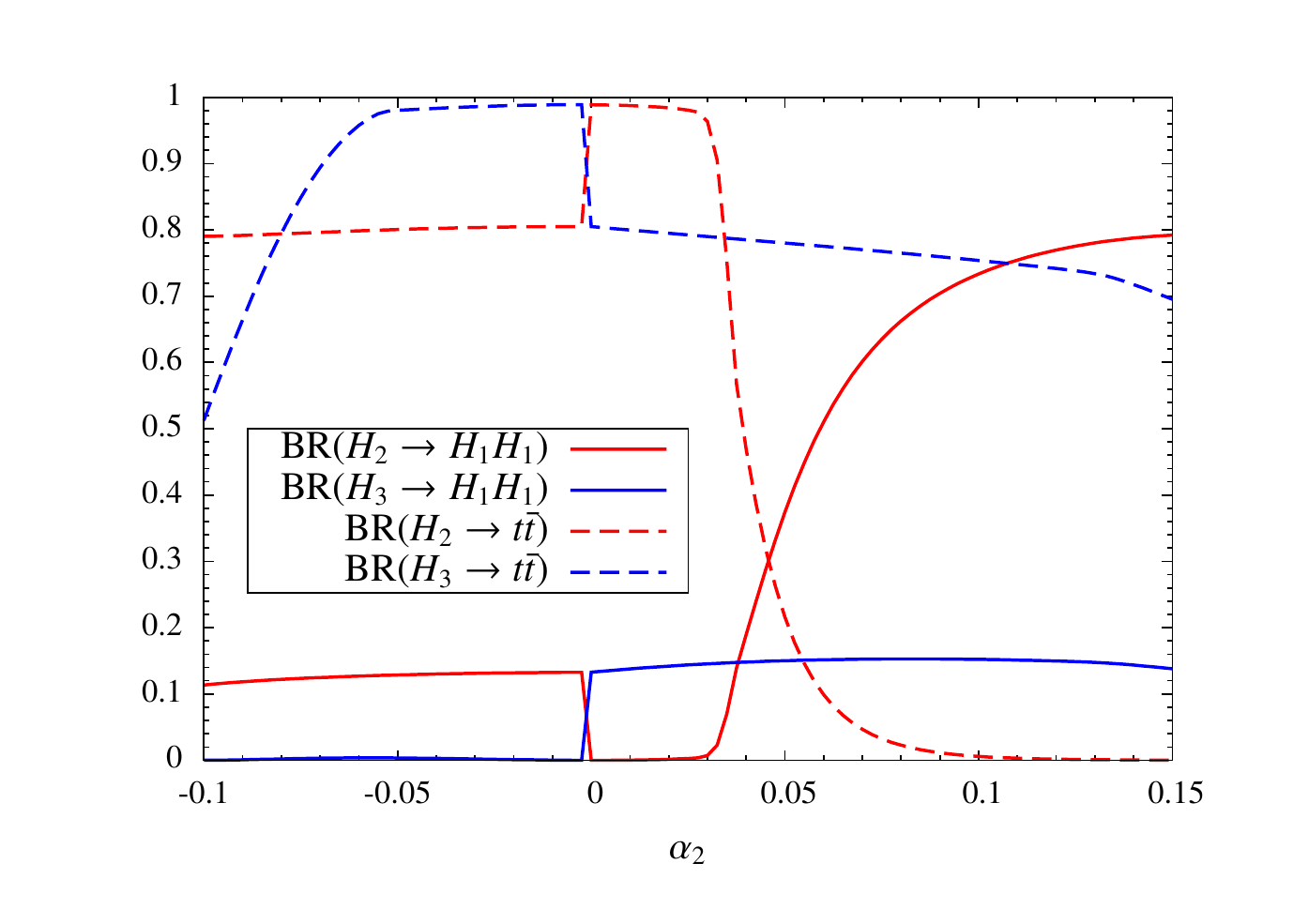}
%\vspace*{-1.3cm}
  \caption{The $H_2$ (red) and $H_3$ (blue) branching ratios into $t\bar{t}$
    (dashed/upper) and $H_1 H_1$ (full/lower). At $\alpha_2=0$, $H_2$ and
    $H_3$ change their role, causing the jump in the branching ratios.
\label{fig:h2h3bran}}
\end{center}
\end{figure}

Figure~\ref{fig:h2h3masses} shows the masses of the next-to-lightest
and heaviest neutral Higgs bosons, $H_2$ and $H_3$, as a function of
$\alpha_2$. At $\alpha_2=0$ there is a cross-over and the Higgs bosons
change their roles: the initially lighter $H_2$ becomes heavier than
$H_3$. Still, we stick to our convention and call the heaviest Higgs
boson $H_3$ and the next heavier one $H_2$. Thus, we have for $-0.13
\le \alpha_2 \le 0.15$, the mass variations
\beq
377.6 \mbox{ GeV } \ge M_{H_2} \ge 277.0 \mbox{ GeV} \quad \mbox{and}
\quad 1398.2 \mbox{ GeV } \ge M_{H_3} \ge 377.6 \mbox{ GeV} \;.
\eeq
Both Higgs bosons are heavy enough to decay on-shell into an $H_1$
pair, so that we can expect the cross section to be larger than in the
SM case. This is confirmed by Fig.~\ref{fig:nloc2hdmcxn}, which shows the cross
section for $H_1 H_1$ production at NLO QCD as a function of
$\alpha_2$ at a c.m.~energy of 14 TeV. 
The smallest value of the cross section is obtained for $\alpha_2 =
-0.13$. With a value of 604.14~fb it exceeds by far the SM cross
section of 38.19~fb.\footnote{This value of the SM cross section
  differs from the one quoted in
  \cite{Borowka:2016ypz}, as we do
  not include the top quark mass effects and work with a different pdf
set.} At $\alpha_2=0.03$, we observe a strong increase
in the cross 
section. The largest value, given at $\alpha_2=0.15$, is 28.48~pb. The strong
increase at $\alpha_2=0.03$ can be understood by inspecting the $H_2$
and $H_3$ branching ratios. They are shown in Fig.~\ref{fig:h2h3bran}
for the decays into $t\bar{t}$ (dashed) and $H_1H_1$ (full) for $H_2$
(red) and $H_3$ (blue). 
The cross-over at $\alpha_2=0$, where $H_2$ becomes heavier
than $H_3$ and they change their roles, is clearly visible by the jump
in the branching ratios. As can be
inferred from the plot, the $H_2$ branching ratio into $H_1 H_1$
strongly increases for $\alpha_2 \ge 0.03$. The $H_2$ mass value here
drops below the $t\bar{t}$ threshold, so that this decay channel gets
closed and the branching ratio into $H_1H_1$ becomes large and even
dominating, as the $H_2$ couplings to the gauge bosons are
suppressed. This increase explains the increase in the Higgs pair
production cross section. Also resonant $H_3$ production with
subsequent decay into $H_1 H_1$ plays a role for positive $\alpha_2$
although it is much less important. At negative $\alpha_2$ only the
$H_2$ branching ratio into $H_1 H_1$ is non-negligible and contributes
to the resonant production.\s

\begin{figure}[t!]
\begin{center}
\includegraphics[scale=0.8]{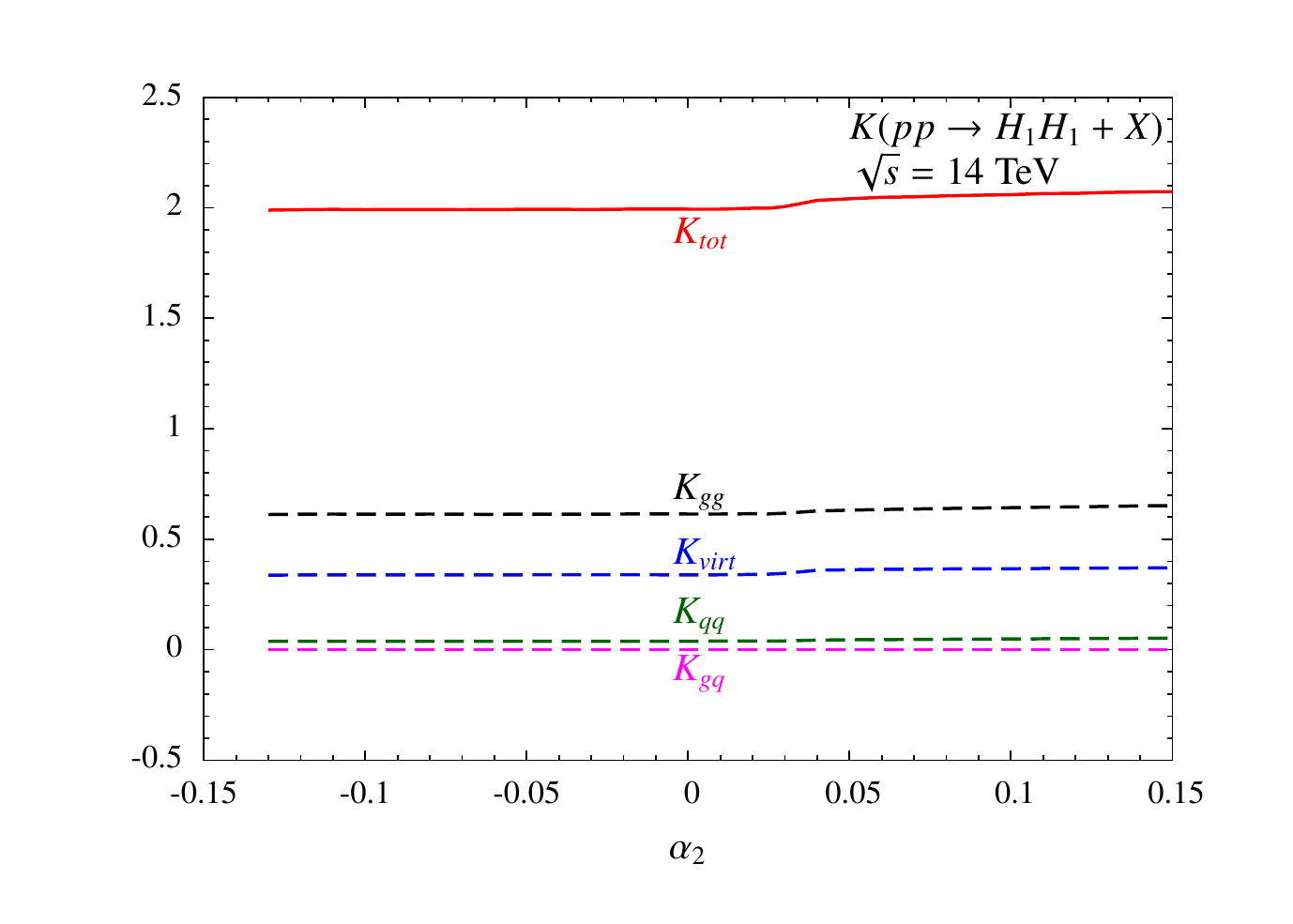}
%\vspace*{-1.3cm}
  \caption{$K$-factors of the QCD-corrected gluon fusion cross section
    $\sigma(pp\to H_1 H_1 + X)$ at the LHC with c.m.~energy
    $\sqrt{s}=14$~TeV. The dashed lines correspond to the individual
    contributions of the QCD corrections, $K_i = \Delta \sigma_i
    /\sigma_{\text{LO}}$ $(i=\text{virt},gg,gq,g\bar{q})$. The C2HDM
    mixing angle has been varied, while all other parameters are given
    as in Eq.~(\ref{eq:scenario}).
\label{fig:c2hdmkfac}} 
\end{center}
\end{figure}

We now turn to the investigation of the $K$-factor, $K =
\sigma_{\text{NLO}}/\sigma_{\text{LO}}$, which is displayed in
Fig.~\ref{fig:c2hdmkfac} together with the individual $K$-factors of
the virtual and real corrections. Again, in the total $K$-factor the
NLO (LO) cross section is evaluated with NLO (LO) parton densities and
$\alpha_s$.
The $K$-factor varies between 1.99 at $\alpha_2=-0.13$ and 2.07 at
$\alpha_2=0.15$. Between $\alpha_2 = 0.03$ and 0.04, where the total
cross section gets strongly enhanced, the $K$-factor
increases a little bit. The maximum deviation from the SM
$K$-factor is found to be
\beq
\delta_{\,\text{max}}^{\,K,\alpha_2} = 0.071
\eeq
for $\alpha_2 = 0.15$. While the deviation in the $K$-factor is small,
the deviation in the absolute cross section is much more
substantial. For $\alpha_2=0.15$ we have
\beq
\delta_{\,\text{max}}^{\,\sigma,\alpha_2}  = 745 \;.
\eeq
This exceeds by far the deviations found in the EFT approach, and is
due to the resonant production of a heavy Higgs boson, subsequently
decaying into $H_1 H_1$. The resonant contribution from $H_2$
production with subsequent decay into $H_1 H_1$ makes up 27.26~pb of the 
total cross section, $\sigma_{\text{NLO}} = 28.47$~pb, and the one of resonant $H_3$
production yields 1.02~pb.\s

\begin{table}[h]
\begin{center}
\begin{tabular}{c|c|c|c} \toprule
final state & $\sigma_{\text{LO}}$ [pb] & $\sigma_{\text{NLO}}$ [pb] &
  $K$-factor\\ \midrule
$H_1 H_1$ & $0.555$ & 1.105 & 1.992 \\
$H_1 H_2$ & $1.939 \cdot 10^{-2}$ & $3.609 \cdot 10^{-2}$ & 1.862 \\
$H_1 H_3$ & $1.153 \cdot 10^{-2}$ & $2.011 \cdot 10^{-2}$ & 1.744\\
$H_2 H_2$ & $1.115 \cdot 10^{-3}$ & $1.948 \cdot 10^{-3}$ & 1.748\\
$H_2 H_3$ & $9.910 \cdot 10^{-4}$ & $1.616 \cdot 10^{-3}$ & 1.631\\
$H_3 H_3$ & $1.172 \cdot 10^{-4}$ & $1.674 \cdot 10^{-4}$ & 1.428 
\\ \bottomrule
\end{tabular}
\caption{The LO and NLO Higgs pair production cross sections and the
  $K$-factor, $K=\sigma_{\text{NLO}}/\sigma_{\text{LO}}$, for the
  final states $H_1 H_1$, $H_1 H_2$, $H_1 H_3$, $H_2 H_2$, $H_2 H_3$ and $H_3
  H_3$. \label{tab:otherhh}}
\end{center}
\end{table}
In Table~\ref{tab:otherhh} we list for our initial scenario defined in
Eq.~(\ref{eq:scenario}) the LO and the NLO cross sections as well as
the total $K$-factor for all final states $H_iH_j$ ($i,j=1,2,3$). 
With increasing mass of the final state Higgs pair the cross sections
decrease as expected. For the Higgs mass
values of our scenario, the only final state that 
can include resonant contributions, is $H_1 H_2$ production: A
resonantly produced $H_3$ subsequently decays into $H_1 H_2$. The
branching ratio $\mbox{BR} (H_3 \to H_1 H_2) = 0.647 
\cdot 10^{-4}$ is, however, very small, so that resonant production
does not play a role in this case.\footnote{The dominant branching ratios are
  those into the $t\bar{t}$ and $ZH_2$ final states.} All non-resonant
cross sections exhibit smaller $K$-factors 
than $H_1 H_1$ production with resonant contributions. They deviate
significantly from the SM $K$-factor 1.94 and lie between 1.43 and
1.86. We furthermore 
observe that the heavier the final state the smaller becomes the
$K$-factor. These findings are in accordance with previous
investigations of the MSSM Higgs sector, where, depending on the final
state, the $K$-factor ranges between about 1.73 and 1.96 \cite{Dawson:1998py}. 

%%%%%%%%%%%%%%%%%%%%%%%%%%%%%%%%%%%%%%%%%%%%%%%%%%%%%%%
\section{Conclusions \label{sec:concl}}
The Higgs sector plays an important role in the search for NP. While
in extensions beyond the SM, the 125~GeV Higgs boson 
needs to have SM-like couplings to the other SM particles, the
trilinear Higgs self-coupling and consequently Higgs pair production can still
deviate significantly from the SM expectations. CP-violation in the
Higgs sector plays an important role to explain the observed
baryon-antibaryon asymmetry. In this work we computed the NLO QCD
corrections to Higgs pair production including CP violation. We worked
in the large top mass limit and performed the calculation on the one
hand in the effective field theory approach, where NP effects are
parametrised by higher-dimensional operators. On the other hand, we
resorted to a specific benchmark model, given by the CP-violating
2HDM. \s

The various contributions to Higgs pair production are affected differently by the
QCD corrections. In the EFT approach including CP-violating effects,
we found that the $K$-factor is 
changed by several per cent only in the investigated parameter regions that
are compatible with the LHC Higgs data. This reflects the dominance of
the soft and collinear gluon effects in the QCD corrections. The
impact of the novel dimension-6 operators on the absolute value of the
cross section is much more important, however, as already found
previously in the case of the CP-conserving EFT. \s

Also in the C2HDM, the $K$-factor of SM-like $H_1 H_1$ production
varies only by a few percent in the investigated parameter range. For
the other possible pair production processes with heavier Higgs
bosons in the final state we find smaller $K$-factors, in accordance
with previous findings in the MSSM, representing a specific
realisation of the 2HDM model. The total cross sections can, however,
be much larger than in the SM. This is due to the possibility of
resonant heavy Higgs production with subsequent decay into the Higgs pair final
state. \s

With $K$-factors between 1.4 and 2.1 in the C2HDM and 1.9 and 2.0 in
the EFT approach, the inclusion of the QCD corrections in the gluon
fusion process is necessary for reliable predictions
of the cross section. 

%%%%%%%%%%%%%%%%%%%%%%%%%%%%%%%%%%%%%%%%%%%%%%%%%%%%%%%
\subsubsection*{Acknowledgments}
RG is supported by a European Union COFUND/Durham Junior Research
Fellowship under the EU grant number 609412. 
We are grateful to Philipp Basler, Marco Sampaio, Rui Santos and Jonas Wittbrodt for
discussions on the C2HDM. 
We would like to thank Jonas Wittbrodt for providing us with viable C2HDM
scenarios.

%%%%%%%%%%%%%%%%%%%%%%%%%%%%%%%%%%%%%%%%%%%%%%%%%%%%%%%%%%
\section*{Appendix}
\begin{appendix}
\section{Gluon Fusion into Higgs Pairs in the SILH
  Approximation \label{app:silhggfus}} 
The SILH approximation for NP effects is valid in case of small shifts $\delta
\bar{c}_i$ in the Higgs couplings $c_i$ away from the SM values
$c_i^{\text{SM}}$, {\it i.e.}
\beq
\mbox{SILH:} \qquad c_i = c_i^{\text{SM}} (1+ \delta \bar{c}_i) \;,
\qquad \mbox{with} \quad 
\delta \bar{c}_i \ll 1 \;.
\eeq
In the non-linear case arbitrary values are allowed for the coupling
coefficients and terms quadratic in $\delta c_i$ have to be
included. This also avoids non-physical observables, as {\it
  e.g.}~negative cross sections. In contrast, in the SILH approach an expansion linear in
$\delta \bar{c}_i$ has to be performed. 
With 
\beq
c_t &=& 1 + \delta \bar{c}_t \equiv 1 - \frac{\bar{c}_H+2
  \bar{c}_u}{2} \;, \quad c_{tt} = \delta \bar{c}_{tt} \equiv  -
\frac{\bar{c}_H + 3 \bar{c}_u }{2} \;, \quad c_3 = 1+ \delta
\bar{c}_3 \equiv 1- \frac{3 \bar{c}_H - 2 \bar{c}_6}{2} \;, \nonumber
\\
c_g &=& \delta \bar{c}_g = \delta \bar{c}_{gg} \equiv \bar{c}_g \left(
  \frac{4 \pi}{\alpha_2} \right) \;, \quad
\tilde{c}_t = \delta \widetilde{\bar{c}}_t \equiv
-\widetilde{\bar{c}}_u \;, \quad \tilde{c}_{tt} 
= \delta \widetilde{\bar{c}}_{tt} \equiv  - 
\frac{3 \widetilde{\bar{c}}_u }{2} \;, \nonumber \\
 \tilde{c}_g &=&
\delta \widetilde{\bar{c}}_g = \delta \widetilde{\bar{c}}_{gg} \equiv 
\widetilde{\bar{c}}_g \left(  \frac{4 \pi}{\alpha_2}
\right) \;, 
\eeq
{\it cf.}~Eq.~(\ref{eq:nonlincoeff}), this yields for the LO partonic cross
section Eq.~(\ref{eq:losigma}) in the SILH parametrisation
\beq
\hat{\sigma}_{\text{LO}}^{\text{SILH}} (gg \to hh) &=&
\hspace*{-0.2cm} \int_{\hat{t}_-}^{\hat{t}_+} 
d\hat{t} \, \frac{G_F^2 \alpha_s^2(\mu_R)}{256 (2\pi)^3} \times 
\label{eq:losigmasilh} \\ && 
\hspace*{-0.2cm} \bigg[
\left| \bar{C}^e_\Delta F_\Delta + F^e_\Box \right|^2 +
| G^e_\Box|^2  + 2\mbox{Re} \bigg\{ \left( \bar{C}_\Delta
  F^e_\Delta + F^e_\Box\right) \, \delta \bar{c}_{tt} \, F^{e*}_\Delta \nonumber \\ 
 && \hspace*{-0.2cm} 
+ \left[ |\bar{C}_\Delta F^e_\Delta|^2 + 3\, 
  \bar{C}_\Delta F^e_\Delta F^{e*}_\Box + 2 \, (|F^e_\Box|^2  +
  |G^e_\Box|^2) \right] \delta \bar{c}_t \nonumber \\
&& \hspace*{-0.2cm}
+ \left(\bar{C}_\Delta F^e_\Delta + F^e_\Box \right)^*
\left[ \bar{C}_\Delta F^e_\Delta \delta \bar{c}_3 + 8 
\left(\bar{C}_\Delta \delta \bar{c}_g + \delta \bar{c}_{gg}
\right) \right] \bigg\} \bigg] \;, \nonumber
\eeq
where
\beq
\bar{C}_\Delta \equiv \lambda_{hhh}^{\text{SM}}
\frac{M_Z^2}{\hat{s}-M_h^2+i M_h \Gamma_h} \;, \qquad \mbox{with}
\qquad
\lambda_{hhh}^{\text{SM}} = \frac{3 M_h^2}{M_Z^2} \;.
\eeq
The NLO SILH cross section is obtained from
Eqs.~(\ref{eq:nlocxn})--(\ref{eq:contrib4}) by replacing 
\beq
\hat{\sigma}_{\text{LO}} \to \hat{\sigma}_{\text{LO}}^{\text{SILH}} \qquad
\mbox{and} \qquad C \to C^{\text{SILH}} \;, 
\eeq
with
\beq
C^{\text{SILH}} &=& \pi^2 + \frac{33-2N_F}{6} \log \frac{\mu_R^2}{Q^2} +
\frac{11}{2} + \left[ \int_{\hat{t}_-}^{\hat{t}_+} d\hat{t} \, 
\tilde{\hat{\sigma}}_{\text{LO}}^{\text{SILH}} \right]^{-1} \times
\nonumber \\
&& 
\mbox{Re} \, \int_{\hat{t}_-}^{\hat{t}_+} d\hat{t}
   \, \bigg\{ \left[a_1 -44 (\bar{C}_\Delta^* \delta \bar{c}_g + 
     \delta \bar{c}_{gg}) \right] (\bar{C}_\Delta F^e_\Delta + F^e_\Box )
+  a_1 \left[  F^e_\Delta \, \delta \bar{c}_{tt} 
\right.
\nonumber  \\
&& \left. 
     + (3
       \bar{C}_\Delta F^e_\Delta + 4  F^e_\Box) \delta \bar{c}_t + 8
       (\bar{C}_\Delta +3  F^e_\Box + 3 \bar{C}_\Delta F^e_\Delta ) \delta
       \bar{c}_g + 8  \delta \bar{c}_{gg} + \bar{C}_\Delta
       F^e_\Delta \delta \bar{c}_3 \right]
\nonumber
\\ &&  + \left[ 1 +4 \, \delta \bar{c}_t + 24 \, \delta \bar{c}_g \right] a_2
    \frac{p_T^2}{2 \hat{t} 
      \hat{u}} (Q^2-2M_h^2) G^e_\Box  \bigg\}
\;, \label{eq:ccoeffsilh}
\eeq
where
\beq
\tilde{\hat{\sigma}}_{\text{LO}}^{\text{SILH}} =
\hat{\sigma}_{\text{LO}}^{\text{SILH}} \left[ \frac{G_F^2 \alpha_s^2
    (\mu_R)}{256 (2\pi)^3} \right]^{-1} \;.
\eeq
As can be inferred from Eqs.~(\ref{eq:losigmasilh}) and (\ref{eq:ccoeffsilh}), in the SILH
approximation inclusive Higgs pair production is not affected by
CP-violating effects at LO in the coupling deviation, {\it i.e.}~at
the dimension-6 level. 
\end{appendix}

\vspace*{0.5cm}
%%%%%%%%%%%%%%%%%%%%%%%%%%%%%%%%%%%%%%%%%%%%%%%%%%%%%%%
%\bibliographystyle{h-physrev}
%\bibliography{cpviolhpairnloeff.bib}   % name your BibTeX data base

\begin{thebibliography}{999}
\bibitem{Aad:2012tfa}
ATLAS Collaboration, G.~Aad {\em et~al.},
 Phys.Lett. {\bf B716}, 1 (2012), arXiv:1207.7214.

\bibitem{Chatrchyan:2012ufa}
CMS Collaboration, S.~Chatrchyan {\em et~al.},
 Phys.Lett. {\bf B716}, 30 (2012), arXiv:1207.7235.

\bibitem{Englert:2014uua}
C.~Englert {\em et~al.},
 J. Phys. {\bf G41}, 113001 (2014), arXiv:1403.7191.

\bibitem{DiLuzio:2017tfn}
L.~Di~Luzio, R.~Grober, and M.~Spannowsky, arXiv:1704.02311.

\bibitem{Dawson:1998py}
S.~Dawson, S.~Dittmaier, and M.~Spira,
 Phys. Rev. {\bf D58}, 115012 (1998), hep-ph/9805244.

\bibitem{Djouadi:1999gv}
A.~Djouadi, W.~Kilian, M.~Muhlleitner, and P.~M. Zerwas,
 Eur. Phys. J. {\bf C10}, 27 (1999), hep-ph/9903229.

\bibitem{Djouadi:1999rca}
A.~Djouadi, W.~Kilian, M.~Muhlleitner, and P.~M. Zerwas,
 Eur. Phys. J. {\bf C10}, 45 (1999), hep-ph/9904287.

\bibitem{Muhlleitner:2000jj}
M.~M. Muhlleitner,
 {\em {Higgs particles in the standard model and supersymmetric
  theories}},
 PhD thesis, Hamburg U., 2000, hep-ph/0008127.

\bibitem{Glover:1987nx}
E.~W.~N. Glover and J.~J. van~der Bij,
 Nucl. Phys. {\bf B309}, 282 (1988).

\bibitem{Plehn:1996wb}
T.~Plehn, M.~Spira, and P.~M. Zerwas,
 Nucl. Phys. {\bf B479}, 46 (1996), hep-ph/9603205,
 [Erratum: Nucl. Phys. {\bf B531}, 655 (1998)].

\bibitem{Baglio:2012np}
J.~Baglio {\em et~al.},
 JHEP {\bf 04}, 151 (2013), arXiv:1212.5581.

\bibitem{Borowka:2016ehy}
S.~Borowka {\em et~al.},
 Phys. Rev. Lett. {\bf 117}, 012001 (2016), 1604.06447,
 [Erratum: Phys. Rev. Lett. {\bf 117} 079901 (2016)].

\bibitem{Borowka:2016ypz}
S.~Borowka {\em et~al.},
 JHEP {\bf 10}, 107 (2016), arXiv:1608.04798.

\bibitem{Heinrich:2017kxx}
G.~Heinrich, S.~P. Jones, M.~Kerner, G.~Luisoni, and E.~Vryonidou, arXiv:1703.09252.

\bibitem{Ellwanger:2013ova}
U.~Ellwanger,
 JHEP {\bf 08}, 077 (2013), arXiv:1306.5541.

\bibitem{Nhung:2013lpa}
D.~T. Nhung, M.~Muhlleitner, J.~Streicher, and K.~Walz,
 JHEP {\bf 11}, 181 (2013), arXiv:1306.3926.

\bibitem{No:2013wsa}
J.~M. No and M.~Ramsey-Musolf,
 Phys. Rev. {\bf D89}, 095031 (2014), arXiv:1310.6035.

\bibitem{Heng:2013cya}
Z.~Heng, L.~Shang, Y.~Zhang, and J.~Zhu,
 JHEP {\bf 02}, 083 (2014), arXiv:1312.4260.

\bibitem{Bhattacherjee:2014bca}
B.~Bhattacherjee and A.~Choudhury,
 Phys. Rev. {\bf D91}, 073015 (2015), arXiv:1407.6866.

\bibitem{King:2014xwa}
S.~F. King, M.~Muhlleitner, R.~Nevzorov, and K.~Walz,
 Phys. Rev. {\bf D90}, 095014 (2014), arXiv:1408.1120.

\bibitem{Chen:2014ask}
C.-Y. Chen, S.~Dawson, and I.~M. Lewis,
 Phys. Rev. {\bf D91}, 035015 (2015), arXiv:1410.5488.

\bibitem{Martin-Lozano:2015dja}
V.~Martin~Lozano, J.~M. Moreno, and C.~B. Park,
 JHEP {\bf 08}, 004 (2015), arXiv.1501.03799.

\bibitem{Wu:2015nba}
L.~Wu, J.~M. Yang, C.-P. Yuan, and M.~Zhang,
 Phys. Lett. {\bf B747}, 378 (2015), arXiv:1504.06932.

\bibitem{He:2015spf}
H.-J. He, J.~Ren, and W.~Yao,
 Phys. Rev. {\bf D93}, 015003 (2016), arXiv:1506.03302.

\bibitem{Dawson:2015haa}
S.~Dawson and I.~M. Lewis,
 Phys. Rev. {\bf D92}, 094023 (2015), arXiv:1508.05397.

\bibitem{Batell:2015koa}
B.~Batell, M.~McCullough, D.~Stolarski, and C.~B. Verhaaren,
 JHEP {\bf 09}, 216 (2015), arXiv:1508.01208.

\bibitem{Zhang:2015mnh}
W.-J. Zhang {\em et~al.},
 Phys. Rev. {\bf D92}, 116005 (2015), arXiv:1512.01766.

\bibitem{Bojarski:2015kra}
F.~Bojarski, G.~Chalons, D.~Lopez-Val, and T.~Robens,
 JHEP {\bf 02}, 147 (2016), arXiv:1511.08120.

\bibitem{Grober:2015cwa}
R.~Grober, M.~Muhlleitner, M.~Spira, and J.~Streicher,
 JHEP {\bf 09}, 092 (2015), arXiv:1504.06577.

\bibitem{Grober:2016wmf}
R.~Grober, M.~Muhlleitner, and M.~Spira,
 JHEP {\bf 06}, 080 (2016), arXiv:1602.05851.

\bibitem{Kanemura:2016tan}
S.~Kanemura, K.~Kaneta, N.~Machida, S.~Odori, and T.~Shindou,
 Phys. Rev. {\bf D94}, 015028 (2016), arXiv:1603.05588.

\bibitem{He:2016sqr}
S.-P. He and S.-h. Zhu,
 Phys. Lett. {\bf B764}, 31 (2017), arXiv:1607.04497.

\bibitem{Krause:2016xku}
M.~Krause, M.~Muhlleitner, R.~Santos, and H.~Ziesche,
 Phys. Rev. {\bf D95}, 075019 (2017), arXiv:1609.04185.

\bibitem{Baglio:2016bop}
J.~Baglio and C.~Weiland,
 JHEP {\bf 04}, 038 (2017), arXiv:1612.06403.

\bibitem{Huang:2017jws}
T.~Huang {\em et~al.}, arXiv:1701.04442.

\bibitem{Nakamura:2017irk}
K.~Nakamura, K.~Nishiwaki, K.-Y. Oda, S.~C. Park, and Y.~Yamamoto, arXiv:1701.06137.
 
\bibitem{Dawson:2012mk}
S.~Dawson, E.~Furlan, and I.~Lewis,
 Phys. Rev. {\bf D87}, 014007 (2013), arXiv:1210.6663.

\bibitem{Grigo:2013rya}
J.~Grigo, J.~Hoff, K.~Melnikov, and M.~Steinhauser,
 Nucl. Phys. {\bf B875}, 1 (2013), arXiv:1305.7340.

\bibitem{Frederix:2014hta}
R.~Frederix {\em et~al.},
 Phys. Lett. {\bf B732}, 142 (2014), arXiv:1401.7340.

\bibitem{Maltoni:2014eza}
F.~Maltoni, E.~Vryonidou, and M.~Zaro,
 JHEP {\bf 11}, 079 (2014), arXiv:1408.6542.

\bibitem{Degrassi:2016vss}
G.~Degrassi, P.~P. Giardino, and R.~Grober,
 Eur. Phys. J. {\bf C76}, 411 (2016), arXiv:1603.00385.

\bibitem{Baur:2002rb}
U.~Baur, T.~Plehn, and D.~L. Rainwater,
 Phys. Rev. Lett. {\bf 89}, 151801 (2002), hep-ph/0206024.

\bibitem{Gillioz:2012se}
M.~Gillioz, R.~Grober, C.~Grojean, M.~Muhlleitner, and E.~Salvioni,
 JHEP {\bf 10}, 004 (2012), arXiv:1206.7120.

\bibitem{deFlorian:2013uza}
D.~de~Florian and J.~Mazzitelli,
 Phys. Lett. {\bf B724}, 306 (2013), arXiv:1305.5206.

\bibitem{deFlorian:2013jea}
D.~de~Florian and J.~Mazzitelli,
 Phys. Rev. Lett. {\bf 111}, 201801 (2013), arXiv:1309.6594.

\bibitem{Grigo:2014jma}
J.~Grigo, K.~Melnikov, and M.~Steinhauser,
 Nucl. Phys. {\bf B888}, 17 (2014), arXiv:1408.2422.

\bibitem{deFlorian:2016uhr}
D.~de~Florian {\em et~al.},
 JHEP {\bf 09}, 151 (2016), arXiv:1606.09519.

\bibitem{Shao:2013bz}
D.~Y. Shao, C.~S. Li, H.~T. Li, and J.~Wang,
 JHEP {\bf 07}, 169 (2013), arXiv:1301.1245.

\bibitem{deFlorian:2015moa}
D.~de~Florian and J.~Mazzitelli,
 JHEP {\bf 09}, 053 (2015), arXiv:1505.07122.

\bibitem{Agostini:2016vze}
A.~Agostini, G.~Degrassi, R.~Grober, and P.~Slavich,
 JHEP {\bf 04}, 106 (2016), arXiv:1601.03671.

\bibitem{Hespel:2014sla}
B.~Hespel, D.~Lopez-Val, and E.~Vryonidou,
 JHEP {\bf 09}, 124 (2014), arXiv:1407.0281.

\bibitem{Contino:2012xk}
R.~Contino {\em et~al.},
 JHEP {\bf 08}, 154 (2012), arXiv:1205.5444.

\bibitem{Goertz:2014qta}
F.~Goertz, A.~Papaefstathiou, L.~L. Yang, and J.~Zurita,
 JHEP {\bf 04}, 167 (2015), arXiv:1410.3471.

\bibitem{Chen:2014xra}
C.-R. Chen and I.~Low,
 Phys. Rev. {\bf D90}, 013018 (2014), arXiv:1405.7040.

\bibitem{Edelhaeuser:2015zra}
L.~Edelhauser, A.~Knochel, and T.~Steeger,
 JHEP {\bf 11}, 062 (2015), arXiv:1503.05078.

\bibitem{Azatov:2015oxa}
A.~Azatov, R.~Contino, G.~Panico, and M.~Son,
 Phys. Rev. {\bf D92}, 035001 (2015), arXiv:1502.00539.

\bibitem{Lu:2015jza}
C.-T. Lu, J.~Chang, K.~Cheung, and J.~S. Lee,
 JHEP {\bf 08}, 133 (2015), arXiv:1505.00957.

\bibitem{Kilian:2017nio}
W.~Kilian, S.~Sun, Q.-S. Yan, X.~Zhao, and Z.~Zhao, arXiv:1702.03554.

\bibitem{deFlorian:2017qfk}
D.~de~Florian, I.~Fabre, and J.~Mazzitelli, arXiv:1704.05700.

\bibitem{Dawson:2017vgm}
S.~Dawson and C.~W.~Murphy, arXiv:1704.07851.

\bibitem{Corbett:2017ieo}
T.~Corbett, A.~Joglekar, H.~L.~Li and J.~H.~Yu, arXiv:1705.02551.

\bibitem{King:2015oxa}
S.~F. King, M.~Muhlleitner, R.~Nevzorov, and K.~Walz,
 Nucl. Phys. {\bf B901}, 526 (2015), arXiv:1508.03255.

\bibitem{Muhlleitner:2017dkd}
M.~Muhlleitner, M.~O.~P. Sampaio, R.~Santos, and J.~Wittbrodt, arXiv:1703.07750.

\bibitem{Sakharov:1967dj}
A.~D. Sakharov,
 Pisma Zh. Eksp. Teor. Fiz. {\bf 5}, 32 (1967),
 [Usp. Fiz. Nauk161,61(1991)].

\bibitem{Contino:2013kra}
R.~Contino, M.~Ghezzi, C.~Grojean, M.~Muhlleitner, and M.~Spira,
 JHEP {\bf 07}, 035 (2013), arXiv:1303.3876.

\bibitem{Cao:2016zob}
Q.-H. Cao, G.~Li, B.~Yan, D.-M. Zhang, and H.~Zhang, arXiv:1611.09336.

\bibitem{Ginzburg:2002wt}
I.~F. Ginzburg, M.~Krawczyk, and P.~Osland, 
 in *Seogwipo 2002, Linear colliders* 90-94, hep-ph/0211371.

\bibitem{Khater:2003wq}
W.~Khater and P.~Osland,
 Nucl. Phys. {\bf B661}, 209 (2003), hep-ph/0302004.

\bibitem{ElKaffas:2006gdt}
A.~W. El~Kaffas, W.~Khater, O.~M. Ogreid, and P.~Osland,
 Nucl. Phys. {\bf B775}, 45 (2007), hep-ph/0605142.

\bibitem{ElKaffas:2007rq}
A.~W. El~Kaffas, P.~Osland, and O.~M. Ogreid,
 Nonlin. Phenom. Complex Syst. {\bf 10}, 347 (2007), hep-ph/0702097.

\bibitem{WahabElKaffas:2007xd}
A.~W. El~Kaffas, P.~Osland, and O.~M. Ogreid,
 Phys. Rev. {\bf D76}, 095001 (2007), arXiv:0706.2997.

\bibitem{Osland:2008aw}
P.~Osland, P.~N. Pandita, and L.~Selbuz,
 Phys. Rev. {\bf D78}, 015003 (2008), arXiv:0802.0060.

\bibitem{Grzadkowski:2009iz}
B.~Grzadkowski and P.~Osland,
 Phys. Rev. {\bf D82}, 125026 (2010), arXiv:0910.4068.

\bibitem{Arhrib:2010ju}
A.~Arhrib, E.~Christova, H.~Eberl, and E.~Ginina,
 JHEP {\bf 04}, 089 (2011), arXiv:1011.6560.

\bibitem{Barroso:2012wz}
A.~Barroso, P.~M. Ferreira, R.~Santos, and J.~P. Silva,
 Phys. Rev. {\bf D86}, 015022 (2012), arXiv:1205.4247.

\bibitem{Fontes:2014xva}
D.~Fontes, J.~C. Romao, and J.~P. Silva,
 JHEP {\bf 12}, 043 (2014), arXiv:1408.2534.

\bibitem{Giudice:2007fh}
G.~F. Giudice, C.~Grojean, A.~Pomarol, and R.~Rattazzi,
 JHEP {\bf 06}, 045 (2007), hep-ph/0703164.

\bibitem{Aad:2015zhl}
ATLAS, CMS, G.~Aad {\em et~al.},
 Phys. Rev. Lett. {\bf 114}, 191803 (2015), arXiv:1503.07589.

\bibitem{Grober:2010yv}
R.~Grober and M.~Muhlleitner,
 JHEP {\bf 06}, 020 (2011), arXiv:1012.1562.

\bibitem{Contino:2010mh}
R.~Contino, C.~Grojean, M.~Moretti, F.~Piccinini, and R.~Rattazzi,
 JHEP {\bf 05}, 089 (2010), arXiv:1002.1011.

\bibitem{Ferreira:2016jea}
F.~Ferreira, B.~Fuks, V.~Sanz, and D.~Sengupta, arXiv:1612.01808.

\bibitem{Graudenz:1992pv}
D.~Graudenz, M.~Spira, and P.~M. Zerwas,
 Phys. Rev. Lett. {\bf 70}, 1372 (1993).

\bibitem{Spira:1993bb}
M.~Spira, A.~Djouadi, D.~Graudenz, and P.~M. Zerwas,
 Phys. Lett. {\bf B318}, 347 (1993).

\bibitem{Spira:1995rr}
M.~Spira, A.~Djouadi, D.~Graudenz, and P.~M. Zerwas,
 Nucl. Phys. {\bf B453}, 17 (1995), hep-ph/9504378.

\bibitem{Kramer:1996iq}
M.~Kramer, E.~Laenen, and M.~Spira,
 Nucl. Phys. {\bf B511}, 523 (1998), hep-ph/9611272.

\bibitem{Harlander:2005rq}
R.~Harlander and P.~Kant,
 JHEP {\bf 12}, 015 (2005), hep-ph/0509189.

\bibitem{Ellis:1975ap}
J.~R. Ellis, M.~K. Gaillard, and D.~V. Nanopoulos,
 Nucl. Phys. {\bf B106}, 292 (1976).

\bibitem{Shifman:1979eb}
M.~A. Shifman, A.~I. Vainshtein, M.~B. Voloshin, and V.~I. Zakharov,
 Sov. J. Nucl. Phys. {\bf 30}, 711 (1979),
 [Yad. Fiz. 30, 1368 (1979)].

\bibitem{Kniehl:1995tn}
B.~A. Kniehl and M.~Spira,
 Z. Phys. {\bf C69}, 77 (1995), hep-ph/9505225.

\bibitem{Altarelli:1977zs}
G.~Altarelli and G.~Parisi,
 Nucl. Phys. {\bf B126}, 298 (1977).

\bibitem{Lee:1973iz}
T.~D. Lee,
 Phys. Rev. {\bf D8}, 1226 (1973).

\bibitem{Branco:2011iw}
G.~C. Branco {\em et~al.},
 Phys. Rept. {\bf 516}, 1 (2012), arXiv:1106.0034.

\bibitem{Lavoura:1994fv} 
  L.~Lavoura and J.~P.~Silva,
  %``Fundamental CP violating quantities in a SU(2) x U(1) model with many Higgs doublets,''
  Phys.\ Rev.\ {\bf D50}, 4619 (1994),
  hep-ph/9404276.

\bibitem{Botella:1994cs}
  F.~J.~Botella and J.~P.~Silva,
  %``Jarlskog - like invariants for theories with scalars and fermions,''
  Phys.\ Rev.\ {\bf D51} (1995) 3870,
  hep-ph/9411288.

\bibitem{Bian:2016awe}
L.~Bian and N.~Chen,
 JHEP {\bf 09}, 069 (2016), arXiv:1607.02703.

\bibitem{Djouadi:1997yw}
A.~Djouadi, J.~Kalinowski, and M.~Spira,
 Comput. Phys. Commun. {\bf 108}, 56 (1998), hep-ph/9704448.

\bibitem{Djouadi:2006bz}
  A.~Djouadi, M.~M.~Muhlleitner and M.~Spira,
  %``Decays of supersymmetric particles: The Program SUSY-HIT (SUspect-SdecaY-Hdecay-InTerface),''
  Acta Phys.\ Polon.\ {\bf B38} (2007) 635, hep-ph/0609292.

%\bibitem{Butterworth:2010ym}
%J.~M. Butterworth {\em et~al.},
% {THE TOOLS AND MONTE CARLO WORKING GROUP Summary Report from the Les
%  Houches 2009 Workshop on TeV Colliders},
% in {\em {Physics at TeV colliders. Proceedings, 6th Workshop,
%  dedicated to Thomas Binoth, Les Houches, France, June 8-26, 2009}}, 2010,
%  arXiv:1003.1643.

\bibitem{hpair}
See M.~Spira's website \texttt{http://tiger.web.psi.ch/proglist.html}.

\bibitem{Dulat:2015mca}
S.~Dulat {\em et~al.},
 Phys. Rev. {\bf D93}, 033006 (2016), arXiv:1506.07443.

\bibitem{Coimbra:2013qq}
R.~Coimbra, M.~O.~P. Sampaio, and R.~Santos,
 Eur. Phys. J. {\bf C73}, 2428 (2013), arXiv:1301.2599.

\bibitem{ScannerS}
R.~Costa, R.~Guedes, M.~O.~P. Sampaio, and R.~Santos,
 \textsc{ScannerS} project, 2014,
 \texttt{http://scanners.hepforge.org}.

\bibitem{Misiak:2017bgg}
M.~Misiak and M.~Steinhauser,
 Eur. Phys. J. {\bf C77}, 201 (2017), arXiv:1702.04571.

\bibitem{Belle:2016ufb}
A.~Abdesselam {\it et al.} [Belle Collaboration],
  %``Measurement of the inclusive $B\to X_{s+d} \gamma$ branching fraction, photon energy spectrum and HQE parameters,''
  arXiv:1608.02344.

\end{thebibliography}

%%%%%%%%%%%%%%%%%%%%%%%%%%%%%%%%%%%%%%%%%%%%%%%%%%%%

\end{document}